\documentclass[twoside,11pt]{article}
\usepackage{epsfig}

\def\Journal#1#2#3#4{{#1} {#2} (#4) #3 }
\def\CMP{{\em Commun. Math. Phys.}}
\def\FBS{{\em Few-Body Systems} }

\def\NPA{{\em Nucl. Phys.} A}

\def\PRO{{\em Prog. Theor. Phys.}}

\def\PLB{{\em Phys. Lett.} B}

\def\PRL{\em Phys. Rev. Lett.}
\def\PREV{\em Phys. Rev.}

\def\PRA{{\em Phys. Rev.} A}

\def\PRC{{\em Phys. Rev.} C}

\def\ZPA{{\em Z. Phys.} A}

\newcommand{\be}{\begin{equation}}
\newcommand{\ee}{\end{equation}}
\newcommand{\bea}{\begin{eqnarray}}
\newcommand{\eea}{\end{eqnarray}}

\begin{document}

\title{ \vspace{1cm}Shell Model in the Complex Energy Plane}
\author{N.~Michel,$^{1,2}$ W. Nazarewicz, $^{3,4,5}$ M. P{\l}oszajczak,$^6$
T.~Vertse,$^{7,8}$ \\
$^1$Department of Physics, Graduate School of Science, Kyoto University, \\
Kyoto 606-8502, Japan\\
$^2$CEA/DSM/IRFU/SPhN Saclay, F-91191 Gif-sur-Yvette, France\\
$^3$Department of Physics and Astronomy, University of Tennessee, \\
 Knoxville, TN 37996, USA\\
$^4$ Physics Division, Oak Ridge National Laboratory, Oak Ridge,
  TN 37831, USA  \\
$^5$Institute of Theoretical Physics, Warsaw University,\\
ul. Ho\.za 69, PL-00681, Warsaw, Poland\\
$^6$ Grand Acc\'el\'erateur National d'Ions Lourds (GANIL),\\
 CEA/DSM -- CNRS/IN2P3, BP 5027, F-14076 Caen Cedex 5, France\\
$^7$Institute of Nuclear Research of the Hungarian Academy of Sciences,\\ 
  H-4001 Debrecen, P. O. Box. 51, Hungary \\
$^8$ University of Debrecen, Faculty of Information Science,\\ H-4010 Debrecen, P. O. Box. 12, Hungary
}
\maketitle
\begin{abstract} 
This work reviews foundations and applications of the  complex-energy
continuum shell model that provides a consistent many-body description
of bound states, resonances, and scattering states. The model  can be
considered a  quasi-stationary open quantum system  extension of the
standard configuration interaction approach for well-bound (closed)
systems. 
\end{abstract}

\clearpage
\newpage
\tableofcontents
\clearpage

\section{Introduction}
Small quantum systems, whose properties are profoundly affected by
environment, i.e., continuum of scattering and decay channels, are
intensely studied in various fields of physics (nuclear physics, atomic
and molecular physics, nanoscience, quantum optics, etc.). These
different open quantum systems (OQS), in spite of their specific
features, have generic properties, which are common to all weakly
bound/unbound systems close to the threshold. While many of these
phenomena have been originally studied in nuclear reactions, it is not
possible to experimentally control the behavior of the nucleus by
varying external parameters as in, e.g., atoms and molecules, quantum
dots, or microwave resonators. 

\begin{figure}[htb]
\centerline{\includegraphics[trim=0cm 0cm 0cm 0cm,width=0.7\textwidth,clip]{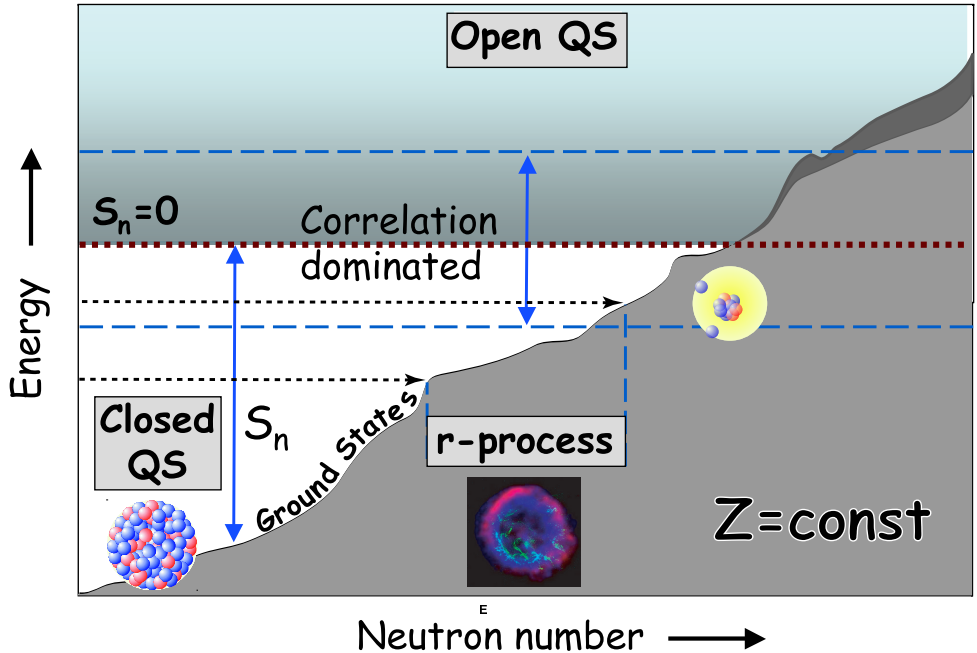}}
\caption{Schematic diagram illustrating various aspects of physics
important in neutron-rich nuclei. One-neutron separation
energies $S_n$, relative to the one-neutron drip-line  limit ($S_n$=0),
are shown for some isotopic chain as a function of $N$.
Weakly bound nuclei, such as halos, inhabit a drip-line region
characterized by very small values of $S_n$.
The unbound
nuclei beyond the one-neutron drip line are resonances. Their widths
(represented by a dark area) vary depending on their excitation energy
and angular momentum. At low excitation energies, well-bound nuclei can
be  considered as closed quantum systems. Weakly bound and unbound
nuclei are open quantum systems that are strongly coupled to the scattering
environment.
The regime of particularly strong many-body correlations 
involving weakly bound and unbound neutrons with 
 $|S_n|$$<$2\,MeV is indicated as  ``correlation dominated" \cite{Dob07}. 
In this region, the neutron chemical potential has the same 
magnitude as the neutron pairing gap. 
The astrophysical r-process is expected to
proceed in the region of low separation
energies, around 2-4 MeV. In this region, effective interactions are
strongly affected by isospin, and the many-body correlations and
continuum effects are essential.}
\label{OQS}
\end{figure}

Nuclear physics contains the core of sub-atomic science with the main focus
on self-organization and stability of nucleonic matter.  Nuclei
themselves are prototypical mesoscopic OQSs and splendid laboratories of
many-body physics. While the number of degrees of freedom in heavy
nuclei is large, it is still very small compared to the number of
electrons in a solid or atoms in a mole of gas. Nevertheless, nuclei
exhibit behaviors that are emergent in nature and present in other
complex systems. Since nuclear properties are profoundly affected by
environment, i.e., the many-body continuum representing scattering and
decay channels, a simultaneous understanding of the structural and
reaction aspects is at the very heart of understanding short-lived
nucleonic matter. 
An essential part of the motion of those exotic systems is in
classically forbidden regions, and their properties are profoundly
impacted by both the continuum and many-body correlations 
(see Fig.~\ref{OQS} and Ref.~\cite{Dob07}).
By studying the limits of nuclear existence, we also
improve our understanding of the ordinary nuclei around us, extending
the nuclear paradigm.  

Resonances are commonly found in various quantum systems, independently
of  their building blocks and the
kinematic regime of their appearance. 
Resonances are genuine intrinsic properties of
quantum systems, associated with their natural frequencies,
and describing preferential decays 
of  unbound states. The
effect of resonances and the non-resonant scattering states can be
considered in the OQS extension of the shell model (SM), the so-called
continuum shell model  (CSM) \cite{[Oko03]}. A particular realization of
the CSM is the  complex-energy CSM based on the Berggren ensemble, the
Gamow Shell Model (GSM). The standard  quantum mechanics  (the Hilbert
space formulation) does not allow the description of state vectors with
exponential growth and exponential decay, such as  resonance states. 
Since the spectrum of an observable is real in the Hilbert space,
the usual procedure for treating unbound resonance  states
is either to extract the
trace of resonances from the real-energy continuum level density or to
describe the resonances by joining the {\em bound} state
solution in the interior region with an asymptotic solution,
e.g.,   within the R-matrix approach \cite{Lane_Thomas,lan}.

These difficulties of the Hilbert space formulation
have been resolved in the Rigged Hilbert Space (RHS)  
\cite{Gel61,Mau68} formulation. Thus, in a broader sense, the mathematical setting
of GSM follows directly from the formulation of quantum mechanics in the
RHS (Gel'fand triple) \cite{Gel61,Mau68} rather than the usual Hilbert
space (see, e.g., Refs.~\cite{Boh78,Mad05}). 
The Gel'fand triple framework allows not only
to formulate rigorously Dirac's formalism of `bras' and `kets'
\cite{Ludwig1,Ludwig2} but also encompasses new concepts, like Gamow
states, and is suitable for extending the domain of quantum mechanics
into the time-asymmetric processes like decays, offering a unified
treatment of  bound, resonance, and scattering states.

In the past,  Gamow states have been discussed in various contexts
\cite{Sie39,Pei59,Hum61,Lin93,Bol96,Fer97,Mad02,Kap03,Her03,Kap05,Jul07,
Mad07}, including those related to the GSM works discussed in this paper.
The subject of this review is to present the foundations and selected
applications of the nuclear SM in the Berggren ensemble. This model
provides a natural generalization of the standard nuclear SM for the
description of configuration mixing in weakly bound states and resonances. 

Alternative  Hilbert-space-based description of the interplay between
many-body scattering states, resonances, and bound states can be
formulated using the concept of projected subspaces
\cite{CSM_Rotter,Phi77}. Realistic studies within the  real-energy CSM
have been presented recently in the framework of  the Shell Model
Embedded in the Continuum (SMEC)  \cite{SMEC,SMEC_2p}. A
phenomenological CSM with an approximate treatment of continuum
couplings has been proposed in Ref. \cite{CSM_Volya}. These models have
the advantage of being able to  provide reaction observables such as
radiative capture and elastic/inelastic
cross sections \cite{SMEC}. However, they rely on a somehow artificial
separation  of the Hilbert space into bound/resonant and non-resonant
scattering parts. This is contrary to GSM, in which  all states are
treated on the same footing. The use of projected subspaces  in SMEC
(CSM) complicates matters significantly when several particles in the
non-resonant scattering continuum are considered. It is only recently
that SMEC has been extended to treat the two-particle continuum
\cite{SMEC_2p} in the context of  two-proton radioactivity
\cite{Bla08}. While numerically demanding, these first applications were
 still subject to several approximations, such as sequential or two-body
cluster decays.

The paper is organized as follows. Section~\ref{GamowS}
discusses the concept of Gamow states and the Berggren basis. The
complex scaling method, employed in GSM to find Gamow states and
regularize integrals, is outlined in Sec.~\ref{complexsc}.
Section~\ref{onebody} describes properties of the one-body Fock space of
the GSM. The many-body GSM is described in Sec.~\ref{GSMS}, and examples
of applications are given in Sec.~\ref{GSMapp}. Finally,
Sec.~\ref{perspectives} outlines perspectives.

\section{Gamow states and the Berggren ensemble}\label{GamowS}

The Gamow states \cite{Gam28,Gur29}
(sometimes called Siegert \cite{Sie39} or  resonant states) 
were introduced for the first time in 1928 by George Gamow to
describe $\alpha$ decay. Gamow introduced  complex-energy eigenstates
\begin{equation}\label{Ecomp}
\tilde{E}_n = E_n - i \frac{\Gamma_n}{2} 
\end{equation}
in order to explain the phenomenon of particle emission in
a quasi-stationary formalism \cite{Baz69}. 
Indeed, if one looks at the temporal part of a decaying
state, which is  $e^{iE_0t/ \hbar} e^{-\Gamma t/(2\hbar)}$, one notices 
that the squared modulus of the wave function has the time-dependence 
$\propto e^{-\Gamma t}$, and one can identify $\Gamma$ with the decay width that defines 
the half-life of the state:
\begin{equation}\label{felez}
T_{1/2}=\frac{\hbar\ln{2}}{\Gamma}.
\end{equation}
The resonant states can be identified with the poles of the scattering matrix $S(k_n)$,
\begin{equation}\label{Epole}
\tilde{E}_n=\frac{\hbar^2}{2m}k^2_n,
\end{equation}
in the complex-momentum plane.  
The bound states are thus resonant states lying on the imaginary-$k$ axis.

In 1968,  Berggren proposed \cite{Ber68} a 
completeness relation for single-particle (s.p.)
resonant states involving a complex-energy 
scattering continuum:
\begin{equation}\label{eq:delb}
\sum_n{u_n(E_n,r)}u_n(E_n,r^{'}) + \int_{L} dE
{u(E,r)}u(E,r^{'})=\delta(r-r^\prime),
\end{equation}
where 
\begin{equation}
u_n(E_n,r)\sim O_l(k_n r)\sim e^{ik_n r}
\end{equation}
and  $k_n =i \kappa_n~~(\kappa_n>0)$ for bound states and 
$k_n= \gamma_n - i \kappa_n~~(\kappa_n,\gamma_n>0)$ for decaying 
resonances in the fourth  quadrant of the complex-$k$ plane.

\begin{figure}[htb]
\centerline{\includegraphics[trim=0cm 0cm 0cm 0cm,width=0.7\textwidth,clip]{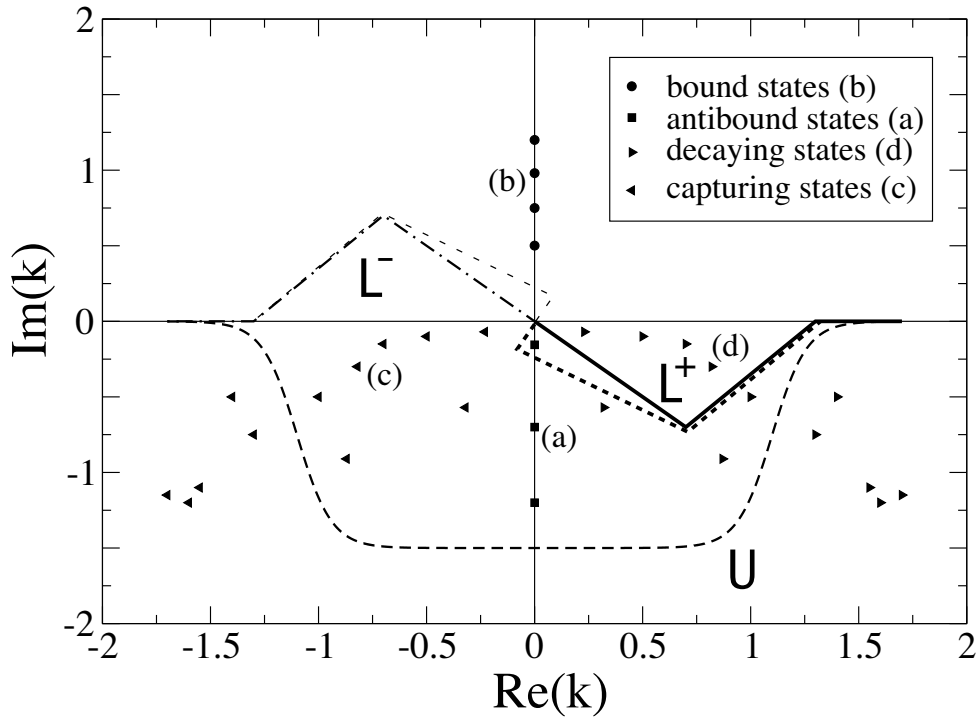}}
\caption{\label{LU_contour}
Location of one-body states  in the  complex momentum plane.
The Berggren completeness relation (\ref{Berggren_comp_rel}) involves
the bound states (b) lying on the imaginary $k$-axis, 
scattering states on the  $L^+$ contour (solid thick line),
and resonant decaying states  (d) in the fourth quarter of the complex-$k$ plane
 lying between the real axis and  $L^+$. 
In the general expansion of the resolvent,  the scattering states
along  the $U$ contour and all
resonant states  lying above  $U$  are included. The antibound
states (a)  can be included in the generalized
completeness relation; in this case the contour $L^+$ has to be slightly
 deformed (dashed thick line).}
\end{figure}

In practical applications, it is more convenient to write (\ref{eq:delb}) 
in  momentum space:
\begin{eqnarray}\label{Berggren_comp_rel}
\sum_{n \in (b,d)} | u_n \rangle \langle u_n | + 
\int_{L^+} |u(k) \rangle \langle u(k)| dk = 1.
\end{eqnarray}
As seen in Fig.~\ref{LU_contour},
bound and quasi-bound states (e.g,  resonances) enter the completeness
relation (\ref{Berggren_comp_rel}) on the same footing. 
The  completeness relation introduced by Berggren reduces to the 
traditional one involving bound and scattering states
if the contour $L^+$ lies on the real $k$-axis.

The radial wave functions corresponding to the $S$-matrix
poles with $Im(k)<0$ diverge 
as $r\rightarrow \infty$; hence,  the
scalar product among basis states has to be
 generalized. Firstly,
a biorthogonal basis
is used for the radial wave function, i.e. a different basis set for
the states in bra and ket  positions. Second, 
radial integrals have to be regularized. These extensions stem from  the fact 
that continuum states belong to RHS. The RHS metric 
 becomes equivalent to the ordinary metric only for
bound states. 

While many important developments took place prior to Berggren's work, e.g., 
analytic properties of the Green's function and the scattering matrix were 
recognized~\cite{Hum61}, and 
Zel'dovich \cite{Zel60} and Hokkyo \cite{Hok65} proposed regularization
methods  to normalize  Gamow vectors; however, the bound and resonant states were 
thought to be non-orthogonal and this caused conceptual difficulties.
In Ref.~\cite{Ber68}, 
Berggren applied  the regularization method of Zel'dovich 
(with a Gaussian convergence factor) to normalize Gamow states and 
proved the completeness relation (\ref{eq:delb}).
%
Another type of regularization was introduced by Romo \cite{Rom68}  who
used the analytic continuation of the norm integral from the upper $k$
halfplane to the resonant state in the lower $k$ halfplane. As an
application, he solved a  two-channel problem involving a Gamow state.
Zim\'anyi introduced yet another way of normalizing   Gamow states by
changing  the strength of the generating potential
\cite{Zim70,Zim70a} and  applying this technique to the coupled Lane
equations for complex-energy isobaric analog resonances.
Bang and Zim\'anyi used a Gamow form
factor in the description of stripping reactions
leading to a  final resonant state \cite{Ban69}.
In 1972,  Gyarmati and Vertse \cite{Gya71} 
proved  the equivalence of normalization procedures of Zel'dovich and
Romo  for neutron resonant states.
They demonstrated the existence of the norm for a proton resonant state
and introduced a new regularization procedure. 
 This {\it external complex scaling} method, applied
in  GSM, is  described in some detail in Sec.~\ref{norm1body}.

\subsection{Antibound states}

Antibound (or virtual) states lie on the
negative  semi-axis of the imaginary $k$: $k_n=-i\kappa_n$~~($\kappa_n>0$)
(see Fig.~\ref{LU_contour}).
They have real and negative energies
that are located in the second Riemann sheet of the complex energy plane
\cite{New82,Nus72,Tay72,Dom81}.
Asymptotically, the radial wave function of a virtual state grows exponentially:
\begin{equation}\label{virt}
w_n(E_n,r)\sim O_l(k_n r)\sim e^{ik_n r}=e^{\kappa_n r}.
\end{equation}
As often discussed in the literature, it is difficult to provide
 a physical
interpretation to virtual states. In the Hilbert space formulation of
quantum mechanics, a virtual state is  not
considered as a  state but as a feature of the system 
(as the second energy sheet is considered unphysical
and inaccessible through direct experiments). In the RHS
formulation, the virtual state can be interpreted both as a vector in
the RHS and as a pole of the $S$-matrix. The latter implies that virtual
states can be ``seen"  only close to the threshold; they manifest
themselves in an
increased localization of low-energy scattering states. Consequently,
the presence of a virtual state at a sufficiently small  energy has an
appreciable influence on the scattering length and the low energy
scattering cross section. Classic examples include   the low-energy 
$\ell$=0 nucleon-nucleon scattering characterized by a large and negative
scattering length \cite{Tay72,Kukulin}. Related to this is an increased
localization of real-energy scattering states just above threshold
\cite{Mig72}.

\begin{figure}[htb]
\centerline{\includegraphics[trim=0cm 0cm 0cm 0cm,width=0.7\textwidth,clip]{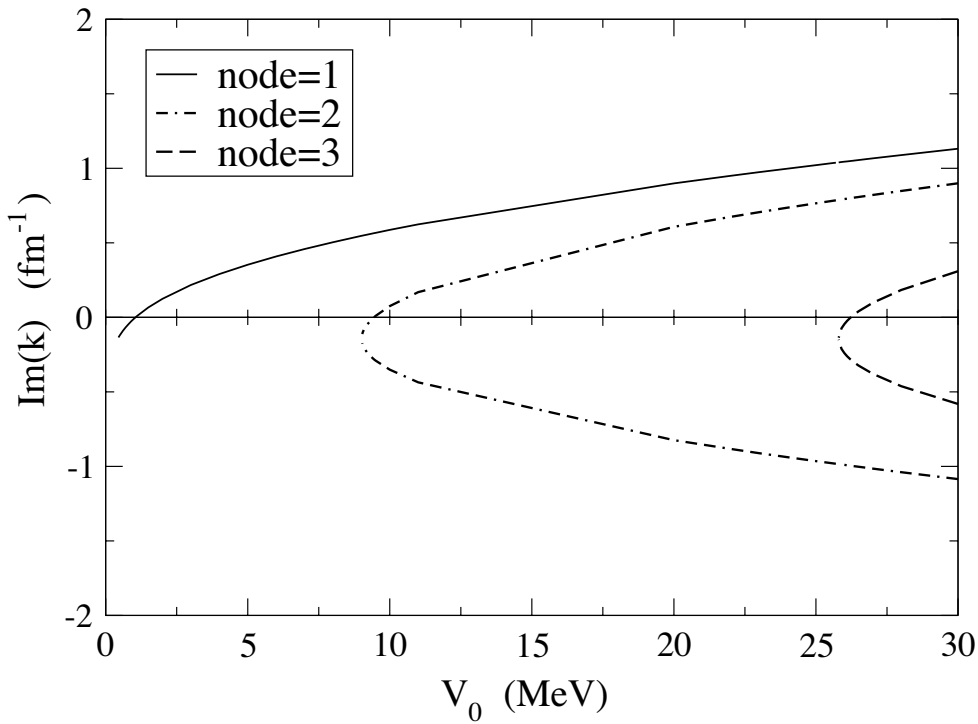}}
\caption{\label{squarewell}
Positions of the bound and antibound neutron $l$=0 
poles with $n$=1,2,3 radial nodes
on the imaginary
 $k$-axis as a function of the  depth $V_0$ of the square well potential
 with radius $R=7$ fm.}
\end{figure}
It is instructive to study the energies of antibound states
as a function of  potential parameters. A detailed study of this problem was done
by Nussenzveig \cite{nuss}  (see also Ref.~\cite{zavin} for a discussion
of the one-dimensional case). To this end, it is convenient to fix the geometry (radius and
diffuseness) of the potential  and follow the trajectories of the bound and
antibound poles as a function of the well depth $V_0>0$. 
If there is no potential barrier, e.g., for $s$-wave  neutrons, no narrow
resonances appear. Here, by increasing $V_0$
 from the minimal value at which the antibound state appears, 
the pole goes across  $k=0$, forming  a bound
state with a radial wave function having one node, $n=1$.  For protons
or for $l>0$ neutrons, a potential barrier is present  which can support
 resonant poles.
In this case, decaying and capturing poles appear in pairs which 
move  towards the imaginary $k$-axis
as $V_0$ increases.  At   certain values of $V_0$  the poles meet at  $k$=0,
  forming  a double singularity. For still larger values of 
$V_0$,  one of the poles becomes a bound
state, while the other one moves down as an antibound state.

 Neutron $l$=0 pole trajectories  are shown in  Fig.~\ref{squarewell} 
 as a function of $V_0$ for a square well potential.
The behavior of the antibound poles conforms to the general  pattern
discussed above. Namely, below $V_0=1.05$ MeV, there appears  only one
antibound state. This pole moves to the upper half $k$-plane and forms a
bound state with one node.  By increasing the depth further from
$V_0=9.05$ MeV, there appear two antibound poles. One of them moves
upwards and at about $V_0=9.5$\,MeV, it emerges as  a $2s$ bound state,
while  the other one moves down and remains antibound. At about
$V_0=26$\,MeV three antibound and two bound states are present. By
increasing $V_0$  to 30 MeV,  the antibound pole lying closest to the
origin becomes bound, resulting in 
 three bound and two antibound
states. At $V_0$=30 MeV,  the $3s$ bound state is closer to the origin
than the corresponding antibound pole. Therefore, there are no antibound
poles between the loosest bound state and the threshold \cite{New82}.
For the cases with  a non-zero barrier, the resonant poles meet in the
origin and one obtains a similar pattern as in   Fig. (\ref{squarewell})
but with the Im($k$)-axis shifted down by about $k=0.15$ fm$^{-1}$. This
shift would change the relative distance  of the antibound and bound
states with respect to the origin, with one antibound pole to appear 
between the loosest bound state and the threshold \cite{New82}.

\subsection{Generalized Berggren representation}

In the original formulation of the completeness relation
 (\ref{Berggren_comp_rel}),
only decaying poles  with 
\begin{equation}
arg(k_n)> -{\pi\over 4}, \quad\quad Re(E_n)>0 
\end{equation}
were considered. Consequently, virtual resonances (resonant poles with
$Re(E_n)<0$) and antibound states were excluded from the Berggren
ensemble. The Berggren representation has been generalized in
Ref.~\cite{Ve89} by including a virtual state into the discrete part of
the basis. The generalization requires a slight deformation of 
$L^+$ as indicated in Fig.~\ref{LU_contour}. More recently, this
generalized Berggren representation was used in a full form (i.e.,
including the scattering component) for the description of drip-line 
 nuclei
$^{11}$Li and $^{72}$Ca \cite{Bet04,Bet05}. A large destructive
interference  between the virtual state and the scattering  contour
states has been  observed. In this picture, the bound ground state of
$^{11}$Li could be explained in terms of  unbound  states of  $^{10}$Li.
 Soon afterwards it was noted \cite{Mic06} that a generalized Berggren
basis is not optimal as far as the number of basis states is
concerned. Indeed, as the halo wave function has a decaying character at
large distances due to the exponentially increasing asymptotics of the
virtual state (\ref{virt}), its inclusion in the basis always induces
strong negative interference with the scattering states. Therefore,
adding antibound states to the basis is not beneficial in GSM
applications, as more
discretized scattering states are necessary to reach a required
precision.

\subsection{Berggren completeness relation for protons}

The  completeness relation by Berggren 
(\ref{Berggren_comp_rel}) can be derived from the
Newton completeness
relation for the set of real-energy eigenstates
of a  short-range
potential \cite{New82}:
\begin{eqnarray}\label{newt}
\sum_{n \in (b)} | u_n \rangle \langle u_n | + 
\int_0^{+\infty} |u(k) \rangle \langle u(k)| dk = 1. \label{Newton_comp_rel}
\end{eqnarray}
The proof
can be carried out  by  deforming the real momentum
axis associated with positive-energy scattering states into the $L^+$
contour of Fig.~\ref{LU_contour} and applying the Cauchy integral
theorem \cite{Lin93}. This is possible because 
resonant states  appear as residues, due to their
$S$-matrix pole character.

Bound and real-energy scattering states 
generated by a potential bearing a Coulomb tail naturally form a
complete set as well, as all self-adjoint operators possess a spectral
decomposition \cite{Dun88}.
However, methods used in Ref.~\cite{Dun88},
 relying on the Lebesgue measure theory, are very general
and fairly abstract. Thus, a simple demonstration 
of the completeness relation for
potentials with a pure Coulomb asymptotic behavior is called for.
The proof of (\ref{newt})  relies on the 
analyticity of the Green's function of a local potential that is 
integrable for $r \rightarrow +\infty$
\cite{New82}. Potentials having a Coulomb tail, proportional to
$r^{-1}$, have not been  considered in Ref.~\cite{New82}. 
Moreover, non-locality often
appears in practical
applications  \cite{Mic04}, so it is important to 
 demonstrate the  completeness relation
for non-local potentials  having a Coulomb asymptotic behavior.
This has been done recently in Refs.~\cite{Mic04,Mic08}, where detailed 
derivations can be found.

In short, the main assumption behind
the  proof presented in  Refs.~\cite{Mic04,Mic08} is that beyond
a finite radius $R_0$ the potential behaves as $r^{-1}$. The corresponding
radial Schr{\"o}dinger equation reads:
\begin{eqnarray}
u''(k,r) = \left( \frac{\ell(\ell+1)}{r^2} + v(r) - k^2 \right) u(k,r) +
\int_{0}^{R_0} w(r,r')~u(k,r')~dr',
\label{Non_local_Schrodinger_eq}
\end{eqnarray}
where $v(r)$ and $w(r,r')$ are respectively the local and non-local 
potentials. For $r > R_0$,
the wave function is a  linear combination
of regular and irregular Coulomb  functions. Their
analytical character and the fact that regular Coulomb wave
functions form a complete set \cite{Mic08,Muk06,Muk78} are important to
the demonstration of the completeness.
Using analytic continuation arguments, one
can also demonstrate  that the set of bound and real-energy
scattering states generated by complex potentials $v(r)$ and $w(r,r')$ 
is complete, provided no exceptional points, i.e., bound states of norm
zero \cite{Hei04}, are present.

\subsection{Foundation of the GSM and interpretation of resonant states}

Heuristic methods often precede a complete formulation of the physical
theory in terms of an adequate mathematical apparatus. This was the case
for Dirac's formulation of quantum mechanics \cite{Dir58} which a
posteriori found a satisfactory setting in the RHS
\cite{Ludwig1,Ludwig2}. This is also the case for theoretical
developments utilizing Gamow states to describe weakly bound and/or
unbound states of quantum systems.  The Berggren completeness relation
\cite{Ber68}, which replaces the real-energy scattering states by the
resonance contribution and a background of complex-energy continuum
states, puts the resonance part of the spectrum on the same footing as
the bound and scattering spectrum. However, resonances do not belong to
the usual Hilbert space, so the mathematical apparatus of 
quantum mechanics in
Hilbert space is inappropriate and cannot encompass concepts such as
Gamow states. Quite unexpectedly, it turned out that the mathematical
structure of the RHS is ready to extend the domain of quantum mechanics
into the time-asymmetric processes such as decays.

The RHS, or Gel'fand triple, is a triad of spaces
\cite{Gel61,Mau68,Boh97}:
\begin{eqnarray}
\Phi \subset {\cal H} \subset \Phi^{\times},
\label{RHS1}
\end{eqnarray}
which represent different completions of the same infinitely dimensional
linear space $\Psi$. $\Phi$ (the subspace of test functions) is a dense
subspace of Hilbert space ${\cal H}$, and $\Phi^{\times}$  (the space of
distributions) is the space of antilinear functionals over $\Phi$.
$\Phi$ is also the largest subspace of  ${\cal H}$ on which expectation
values, uncertainties and commutation relations can be correctly defined
for unbounded operators.

The linear functionals over $\Phi$ are contained in the space $\Phi^{'}$
which is related to another RHS:
 \begin{eqnarray}
 \Phi \subset {\cal H} \subset \Phi^{'}~~.
 \label{RHS2}
 \end{eqnarray}
One should stress that ${\cal H}$  in (\ref{RHS1}) and (\ref{RHS2})
does not have any particular significance in the RHS formalism. For all
physics problems, one needs only the dual pair of spaces $\Phi \subset
\Phi^{\times}$ which characterize the considered quantum system. For
example,  bras and kets, which are related to the continuous part of the
spectrum of an observable, belong to $\Phi^{'}$ and $\Phi^{\times}$,
respectively, and not to ${\cal H}$. Apparently, the RHS provides also a
more convenient framework than ${\cal H}$ to capture physical principles
of quantum mechanics. The Heisenberg's uncertainty relations are
properly defined on $\Phi$, and not on ${\cal H}$.

Besides implementation of Dirac's formalism of bras and kets, the RHS
provides a framework for a quantum mechanical description of common
irreversible processes, like the formation or decay of quantum states.
Indeed, $\Phi^{\times}$ may contain generalized eigenvectors of the
observable(s) with complex eigenvalues. The generalized eigenvectors of
the Hermitian Hamiltonian are  Gamow vectors. 
Examples of the RHS for the observables
having a continuous part in the spectrum can be found, for example, in
Refs. \cite{Civ04,Mad07}.

Gamow vectors are state vectors of resonances. They belong to RHS and
not to ${\cal H}$ since the self-adjoint operators in ${\cal H}$ can
only have real eigenvalues. Like the plane waves (the Dirac kets), the
Gamow functions are not square integrable and must be treated as
distributions. In this way, one succeeds to generate Gamow bras and kets
which fit naturally  in the RHS.  However, unlike the Dirac kets, the
probability density for Gamow functions is not constant but increases
exponentially in space.

Rules for the normalization of Gamow functions and corresponding
completeness relations have been proposed
\cite{Hok65,Ber68,Rom68,Gar76,Zel61,Gya71} independently of the RHS
formulation. Similarly, the momentum space representation for Gamow
functions has been introduced \cite{Her84,Mon91}. RHS offers a
mathematical framework to formulate those results rigorously. In
particular, the RHS provides a unifying dual description of  bound,
resonant and scattering states both in terms of the $S$-matrix and in
terms of the vector in the RHS. 
The transition amplitude
${\cal A}(E_n\rightarrow E)$ from a resonance of energy $E_n$ to a
scattering state at an energy $E$ ($E\geq 0$)\cite{Mad07}:
\begin{eqnarray}
{\cal A}(E_n\rightarrow E)=i\sqrt{2\pi}N_n\delta(E-E_n),
\end{eqnarray}
where $N_n$ denotes the normalization factor, is proportional to the
complex $\delta$-function \cite{complex_distr} which,  far off the threshold energy in the
region of the resonance, can be approximated by the Breit-Wigner
amplitude. This physically expected result provides a formal link
between Gamow states and nearly-Lorentzian  peaks in 
cross-sections for narrow, isolated resonances.

Interpretation of certain aspects of the RHS formulation of quantum
mechanics is still debated, and connections between  Berggren and RHS
formulations continue to be studied (for a recent review, see
\cite{Civ04}). One aspect of this discussion concerns complex matrix
elements of the operators and probabilistic interpretation of the
resonant wave function. As an illustration of this problem for
resonances, let us consider the solution of the  Schr\"odinger equation
$\chi(\mathbf{r},t)$ for a s.p.~Hamiltonian $\hat h$:
\begin{eqnarray}
\label{idofugg}
i\hbar{\partial\over{\partial t}}\chi(\mathbf{r},t)={\hat h}\chi(\mathbf{r},t).
\end{eqnarray}
In the stationary picture, the resonant wave function is the product of time-dependent $\tau(t)$ and
 space-dependent $\psi(\mathbf{r})$ factors:
\begin{eqnarray}
\chi(\mathbf{r},t)=\tau(t) \psi(\mathbf{r}),
\end{eqnarray}
where the time-dependent factor is:
\begin{eqnarray}
\tau(t)=e^{-i{{ \tilde{E}_n}\over \hbar}t}.
\end{eqnarray}
The complex energy $\tilde{E}_n$ (\ref{Ecomp})
 is the generalized eigenvalue of $\hat h$
with the complex wave number $k_n$ (\ref{Epole}).
The corresponding space-dependent eigenfunction $\psi(\mathbf{r},k_n)$,
\begin{equation}
\label{sphwfn}
\psi_n = \psi_{n{\ell}jm} (\mathbf{r},k_n) =  {u_{n{\ell}j}(k_n,r)\over r}
\bigl [Y_{\ell}( \hat r )\chi_s \bigr]_{jm},
\end{equation} 
satisfies an outgoing wave boundary condition. 
At large distances, 
the radial part of the $\ell$=0 wave function
is proportional to $\exp(ik_nr)$. For a
decaying resonance,  $k_n=\gamma_n -i \kappa_n$ ($\gamma_n, \kappa_n>0$),
the time-dependent factor
\begin{eqnarray}
\label{idogamd}
\tau(t)=e^{-i{{\tilde{E}_n}\over\hbar}t}=e^{-i{{E_n}\over\hbar}t}
e^{-{\Gamma_n\over{2\hbar}}t}~~,
\end{eqnarray}
describes an exponential decay with a decay width
$\Gamma_n$. As $r$ increases, the radial part of $\psi(\mathbf{r},k_n)$
oscillates with an exponentially growing amplitude:
\begin{eqnarray}
\label{kifutasz}
e^{ik_n r}=e^{i\gamma_n r}e^{\kappa_n r}=[\cos (\gamma_n r) + i \sin (\gamma_n r)]e^{\kappa_n r}~~.
\end{eqnarray}
The decay process in this picture is unlimited in both time ($t\in
[0,\infty]$) and space ($r\in [0,\infty]$) and, henceforth,  continues
forever. This feature guarantees, in fact, the
particle-number conservation \cite{sasada}: the exponential
temporal decrease of the wave function amplitude is complemented by its
exponential spatial increase, and the divergence of the resonance wave
function assures  that the particle number is conserved.

In general, the quasi-stationary description of a time-dependent
physical process works well when the decaying state is narrow, as in
this case the formation of a state and its subsequent decay can be well
separated in time. A narrow Gamow resonance has a large overlap with a
properly shaped wave packet \cite{Rom84} and its radial wave function is
similar to a bound state wave function  while  the imaginary part is
small inside the nucleus. In the reaction cross section, such a  narrow
resonance shows up as a sharp peak which can be well separated from the
non-resonant background.

Another aspect of the debate on the physical interpretation
of Gamow vectors concerns the expectation value of an operator in a resonant state.
Berggren's interpretation of real and imaginary parts of the expectation
value of an operator is based on the discussion of interference effects
in the reaction cross-sections \cite{Ber78}. Berggren showed that the
imaginary part of the complex cross section calculated with the Gamow
resonance as a final state describes the interference of that resonance
with the non-resonant background. In a later paper
\cite{Ber96}, Berggren demonstrated  that for any operator which commutes with the Hamiltonian,
the expectation value of that operator is the real part of its matrix
element, and the negative of the square of the imaginary part of the
matrix element  can be associated with 
the square of an uncertainty of the expectation
value. This proof holds only for operators which commute with the
Hamiltonian. It turns out that the  Berggren interpretation of the
expectation value \cite{Ber96} is  largely equivalent to  the RHS
formulation by Bohm and Gadella \cite{Boh89} and coincides with it  in
the leading order  \cite{Bol96, Civ99}.

The meaning of a complex radial expectation value was studied in
Ref.~\cite{Bur96} using a model of two coupled channels \cite{Fri91}. It
was found that the radial expectation value in this model can be
interpreted according to Berggren's prescription and that it can be
used to characterize the resonances as long as the wavelength of the
decaying wave is shorter than the radial extension of the resonant state.

\subsection{Applications of single-particle 
resonant states to various nuclear structure problems}

Already in 1970, in the pioneering work employing resonant states, a
truncated resonant basis composed of bound neutron and  resonant proton
states was used to describe isobaric analog resonances 
\cite{Zim70a}. The pole expansion of the radial wave function, the
Green's function, and the scattering matrix were subsequently discussed in
a rigorous way in Ref.~\cite{Ban78} by using the Mittag-Leffler (ML)
expansion. They used the normalization of the pole solutions  introduced
in~\cite{Hok65} and later employed in, e.g.,
Refs.~\cite{Gar76,Cal86}. The authors of Ref.~\cite{Ban78}
suggested that the  pole expansion could be used in (shell-model-type)
nuclear structure calculations; however, they worried about the
non-orthogonality of the basis. In a numerical study~\cite{Ver91},
the usefulness of the ML-type pole expansion of the Green's function was
examined. The question of overcompleteness was finally put to rest
by Berggren and Lind \cite{Ber93}.

Resonant state expansions were
 also used by atomic physicists for calculating the
$S$-matrix,  expressed as a sum of pole terms and a contour
integral~\cite{Alf04}.  The positions of the poles were calculated by using 
the uniform complex scaling (see Sec.~\ref{complexsc}),
while the residues were obtained by using 
 numerical methods described in Refs. \cite{Rit87,Kry89}. 
Neutron emission from a heavy-ion reaction was modeled in a time-dependent
 two-center shell model in Ref.~\cite{Mil86}. Therein, resonant
states  were calculated in the momentum space using the potential
separable expansion (PSE) method \cite{Gya79,Gar78}, in which the potential is
expanded in a square integrable (harmonic oscillator; HO) basis, and the
proper asymptotic of the wave function is guaranteed by the presence of
the free Green's function.

A less severe truncation of the Berggren basis was employed in the late
1980s in the resonant approach to the random phase approximation (RRPA),
in which the contribution from the non-resonant  continuum was neglected
\cite{Ver88,Cur89}. Such an expansion  is usually referred to as a 
 {\it pole approximation}. (A detailed summary of the early theoretical
approaches employing resonant states can be found in Ref.~\cite
{Ver87}.) A similar pole approximation  was used in the
RRPA description of giant resonance escape widths \cite{Dus92}. 
A careful test of 
different pole expansions was carried out in Ref.~\cite{Ver91} by
using a benchmark continuum RPA result for the particle-hole response
function (see Sec.~\ref{CRPA} for details). It was found that the 
ML-type pole expansion was
closer to the continuum RPA, based on the  Green's function, than the pole
approximation. This result was  explained by 
Berggren and Lind~\cite{Ber93} who identified the  
ML-type pole expansion of the resolvent by using 
the so-called $U$ contour (see
Fig.~\ref{LU_contour}) in the complex $k$-plane. Unfortunately,
an expansion using the $U$ contour,
\begin{equation}
\label{Mitexp}
1={1\over2}\sum_{i=a,b,c,d} |u_i\rangle\langle\tilde u_i| + 
{{1}\over{\pi}}\int_{U}|u(k)\rangle dk
\langle u(k^*)|,
\end{equation}
cannot be used in shell-model-like  calculations
because  the factor $1/2$  in the front of the sum over the
resonant  (bound, antibound, capturing, and decaying) states 
destroys the idempotency of the unity operator \cite{Lin94}.
In addition, pole
expansions of type  (\ref{Mitexp}) are
accurate only if very many  poles (up to  some large  energy cut-off) are
included in the sum; 
contributions from broad and virtual resonances
(having  negative real energies) are in fact essential \cite{Lin94}.

In Ref.~\cite{Tol98}, Gamow states were expanded in a finite $r\in[0,R_o]$
interval by using a finite number of square integrable basis states,
called  {\it Siegert pseudostates} (SPS). The expansion coefficients
were determined by solving a normal eigenvalue
problem where the dimension of the system has been doubled. This doubled
system represents an eigenvalue problem with a weight matrix of the
Bloch operator. The weight matrix allows the continuation of the SPS into the
tail region and defines a special inner product in which a
completeness relation can be constructed for the SPS. 
A pole expansion of the Green's function, similar 
to the ML expansion of Ref.~\cite{Lin94}, was carried out. This expansion
was further  exploited in Ref.~\cite{Toy05}. The usefulness of SPS was further
demonstrated in studies of time evolution of an infinitely extended system~\cite{Toy01}.

The resonant and continuum components of the strength function were
calculated by using the Berggren completeness relation in studies of
Coulomb breakup of  $^{11}$Be \cite{Myo98}.
They found the non-resonant contribution of the part of
the contour which returns to the real $k$-axis negligible.  A similar
result was obtained in Ref.~\cite{Hag06}.

In some of the pole expansions mentioned above, complex eigenvalues
and radial wave functions of resonant  states were calculated  using the
 code {\sc gamow} \cite{Ver82}. Later, more  efficient  numerical
techniques have been used based on the  piecewise perturbation method
\cite{Ixa84,Ixa95}. In other work \cite{Zha92} spherical resonant
states were obtained by solving a non-linear set of equations. Bound
and resonant state energies were calculated in code {\sc pseudo} by
using the PSE method in \cite{Kru85} for an axially symmetric potential
by finding the zeroes of the Fredholm determinant of the 
Lippmann-Schwinger equation. Recently a method  was developed in which
the momentum-space Schr\"odinger equation was solved by a contour
deformation method \cite{Hag06}.

More recent applications of pole expansions include a description of 
giant multipole resonances \cite{Ver95} (see Sec.~\ref{CRPA}), and
partial decay widths corresponding to the proton decay 
from the Gamow-Teller and isobaric analog state resonances \cite{Blo96}.
The Berggren representation in its  full form, including the complex scattering 
continuum, was employed in
Ref.~\cite{Lin96} to describe the eigenstates of a realistic 
one-body nuclear Hamiltonian. In this work, eigenstates of a  spherical
symmetric potential  were expanded in a full
Berggren basis of a different potential.
(Examples of such expansions are presented in Sec.~\ref{one_body_calc}.)
The full Berggren ensemble was employed in studies of the average
s.p.~level density \cite{San97} and shell corrections \cite{Ver98,Ver00}. 
Quasiparticle resonances can be constructed from the Gamow resonances in
the BCS approximation by generalizing the standard BCS to the complex
energy plane if the contour integral is neglected. If the non-resonant
contribution is included,  the contribution from the continuum level
density has to be added \cite{Dus07} when calculating the particle
number. However, a  proper description of pairing correlations in weakly bound
systems must go beyond the BCS approximation \cite{Dob96}. 
Here, the tool of choice is the Gamow-HFB method to be mentioned in Sec~\ref{GHF_section}.

Gamow states and resonant  expansions have been extensively used in the
description of alpha, proton, and neutron emission from nuclei. For very
narrow alpha and proton resonances, a single-channel  Gamow
approximation that ignores the non-resonant background is fully adequate
\cite{Del00,Bia01}. An attempt was made to calculate resonant states of
a deformed potential by means of an expansion  in a spherical Berggren
basis \cite{Mag98,Mag99}. However, for a description of narrow  proton
resonances in   deformed nuclei, the accuracy of spherical expansion
turned out to be not satisfactory and alternative approaches were used.
One of them was the direct numerical integration of the set of  coupled
differential equations using the piecewise perturbation method. The
resulting code {\sc ccgamow} was employed  in Ref.~\cite{Ryk99} to explain full
and partial decay widths of protons from axially symmetric
nuclei within an adiabatic approach that ignores the
rotational excitations of the daughter system. A full non-adiabatic
description of proton emitters was carried out in
Refs.~\cite{Kru00,Bar00} using a modified version of the coupled
channels program {\sc nonadi}, and the extension to the triaxial case can be
found in Ref.~\cite{Kru04}. A shell-model analysis of the proton
emission from $^{31}$Cl using Gamow wave functions was performed in
Ref.~\cite{Mar01}. In a study of Ref.~\cite{Gur04}, Gamow states were
used to benchmark  a modified two-potential approach to tunneling problems.

The  applications of resonant states  mentioned so far were concerned
with the s.p.~space. By 2003, however, it became possible to
apply  the  Berggren ensemble to  a general many-body case. In two
parallel studies \cite{Bet02,Mic02}, based on a configuration mixing
approach employing the one-body Berggren ensemble, two-particle
resonances in $^{80}$Ni and  $^{102}$Te were calculated using separable
multipole interactions \cite{Bet02}, and   systematic calculations for
the  neutron-rich nuclei $^{6-10}$He and $^{18-22}$O were carried out
using the surface-delta interaction \cite{Mic03}. More applications soon
followed (see Sec.~\ref{GSMapp}).
 In the following, we shall refer to these two realizations of
the  SM in the complex energy plane as the Gamow Shell Model since the
underlying concept is the same.
 
\subsection{Example 1: resonant state expansions and continuum RPA}\label{CRPA}

Continuum RPA (CRPA) is a useful tool for testing the usefulness and
accuracy of resonant state expansions. For separable multipole-multipole
interactions, the solution of the CRPA equations is obtained  by finding
complex roots of  the corresponding  dispersion
relation~\cite{Ver91,Ver95}. Complex energy poles of the particle-hole
response $R(E)$ yield positions and full widths of the correlated
resonances while the  partial decay widths of these states can be
obtained from the associated residua.

Within CRPA, the particle Green's function in a selected partial wave is 
\begin{equation}
\label{stgreen}
g(r,r^\prime,k)= -{{u(k,r_<)~v(k,r_>)}\over{W(u,v)}},
\end{equation}
where $W$ denotes the Wronskian of the regular $u(k,r)$ and irregular $v(k,r)$ 
solutions
at $r_<$=$\min(r,r^\prime)$ and $r_>$=$\max(r,r^\prime)$,
respectively, at real values of $k$.	
In the standard Berggren expansion using
 the $L$ contour of Fig.~\ref{LU_contour},
the corresponding Green's function can be written as:
\begin{equation}
\label{Bggreen}
g_L(r,r^\prime,k)=
\sum_{i=b,d}{{u_i(k_i,r)~u_i(k_i,r^\prime)}\over{k^2-k_i^2}}+{1\over \pi}
\int_{L} {{u(q,r) u(q,r^\prime)}\over{q(k-q)}} dq.
\end{equation}
On the other hand, if one uses
 the $U$ contour of Fig.~\ref{LU_contour}, the Green's function takes the form:
\begin{equation}
\label{Mlgreen}
g_U(r,r^\prime,k)=
\sum_{i=a,b,d,c}{{u_i(k_i,r)~u_i(k_i,r^\prime)}\over{2k_i(k-k_i})}+{1\over \pi}
\int_{U} {{u(q,r) u(q,r^\prime)}\over{q(k-q)}} dq~.
\end{equation}
The two complex-energy representations of Green's function given by 
 Eqs.~(\ref{Bggreen},\ref{Mlgreen}) and the real-energy
 expression (\ref{stgreen}) are
mathematically equivalent; hence,  they
should give the same results in practical applications. 
 
From a practical point of view,  it is interesting to know
 if one can choose the
complex contours in such a way  that the contributions from
the integrals in
Eqs.~(\ref{Bggreen}) and (\ref{Mlgreen}) become negligible. 
This has been checked in  Ref.~\cite{Ver91}, in which the
particle-hole response functions were calculated for a square-well potential
supplemented by the Coulomb field.
For narrow
and isolated resonances, the partial widths obtained by using the leading 
pole term in 
Eqs.~(\ref{Bggreen}) and (\ref{Mlgreen}) agree with the exact result
within 10\,\%. The
subsequent  calculations of Refs.~\cite{Lin94,Ver95} using the
Woods-Saxon potential demonstrated  that the contribution from the $U$
contour to the p-h response function could be further reduced  but this
does not hold for the $L$ contour. Furthermore, it was shown
\cite{Ber93,Lin93} that neglecting the integral in Eq.~(\ref{Bggreen}) 
destroys the symmetry of the resolvent and causes an artificial
threshold behavior of the Green's function.

\subsection{Example 2: GSM for isobaric analogue resonances in Lane model}\label{IAR}

The isobaric analogue resonance (IAR)  can be
described phenomenologically by the  coupled-channel Lane equations (CCLE) 
 \cite{lan}. The IAR-CCLE problem
offers an excellent opportunity to compare  the complex-energy  GSM approach
with the standard CCLE solution along the real energy axis.
Such a test has been  carried out in Ref.~\cite{Bet08} for the Lane
Hamiltonian. 
Within CCLE, the multichannel Schr\"odinger equation
has been solved by a direct numerical
integration. The resulting proton scattering
$S$-matrix 
has been calculated 
 along the real energy axis in the region of the
IAR and then fitted by using  a single pole approximation,
\begin{equation}
\label{poleform}
S(E_p)=e^{2i\delta_p(E_p)}\left(1-i\frac{\Gamma}{E_p-
\cal{E}_{IAR}}\right),
\end{equation}
to determine
the position $E_r$ and width $\Gamma$ of the IAR.
In Eq.~(\ref{poleform}), 
the background phase shift $\delta_p(E_p)$ was assumed to have
 a linear energy dependence in order
to better reproduce the non-resonant background. 
The best fit parameter values,  denoted  by $E_r(CCLE)$ and $\Gamma(CCLE)$, are
listed in Table \ref{compare}. They coincide with the GSM eigenvalue 
${\cal E}_{IAR}=E_r(GSM)-i{\frac{\Gamma(GSM)}{2}}$ 
 within $1$ keV  even for
the broad states. 
\begin{table}[ht]
\begin{center}
\begin{tabular}{|c|cc|cc|}
\hline
$l~j$&$E_r (GSM)$&$E_r (CCLE)$&$\Gamma (GSM)$&$\Gamma (CCLE)$\\
\hline
$d_{5/2}$&16.445&16.444&0.141&0.140\\
$s_{1/2}$&16.918&16.917&0.156&0.156\\
$d_{3/2}$&17.441&17.440&0.144&0.145\\
\hline
\end{tabular}
\end{center}
\caption{Comparison of the IAR parameters calculated by using  CCLE and GSM 
 for three  partial waves. All energies are in MeV.
\label{compare}}
\end{table}
The integrated effect of the proton continuum along the complex path is small but
essential to yield the correct value of $\Gamma (GSM)$.

\section{Complex scaling method}\label{complexsc}

Another approach  that employs the concept of
complex energies and distorted contour  is the complex scaling method (CS).
Some  aspects of the  CS formalism, e.g., the use
of complex coordinates,  are very relevant to
the GSM. For a comprehensive review of the CS method, we refer the reader
to Reinhardt \cite{Rein82}. Below only the most essential facts are summarized.

\begin{figure}[htb]
\centerline{\includegraphics[trim=0cm 0cm 0cm 0cm,width=0.7\textwidth,clip]
{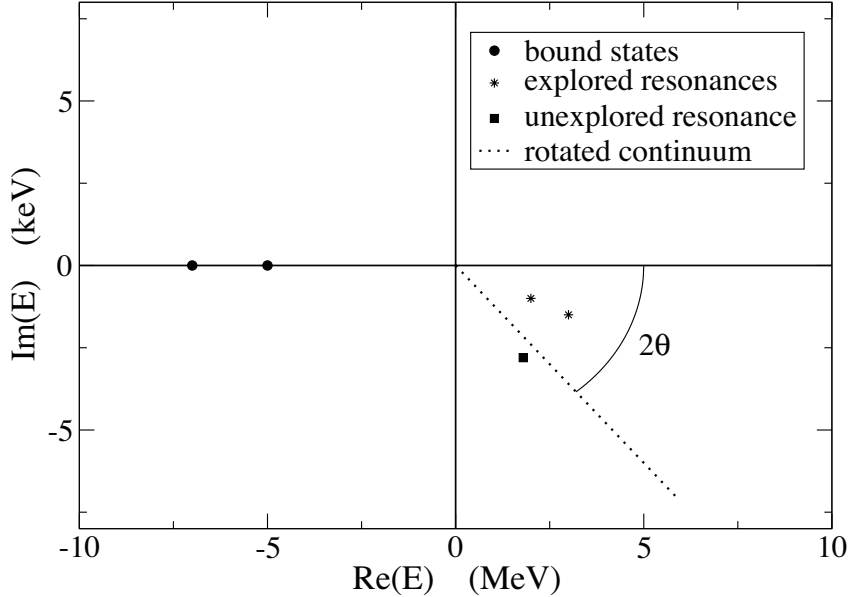}}
\caption{
The complex energy plane showing
bound poles (dots), resonant poles exposed by the complex scaling
transformation (asterisks), the rotated continuum of the scattering
states (dotted line), and a hidden pole (marked by a square; not
exposed  by the  CS transformation).
According to the ABC theorem,  the eigenfunction of the 
exposed poles are  square integrable.
}
\label{cspoles}
\end{figure}

Within CS,  like in GSM,  the resonances are  represented
by poles of the scattering matrix $S(k)$. 
By making use of the CS transformation on particle coordinates,
\begin{equation}
\mathbf{r} \rightarrow  e^{i\theta}{\mathbf{r}},
\end{equation}
one guarantees that 
 wave functions of the selected resonances
become  square integrable
\cite{Agu71,Bal71,Sim72}.
A unitary CS operator   ${\hat U}(\theta)$ acting on
a s.p.~wave function  
gives:
\begin{equation}
{\hat U}(\theta)\psi(\mathbf{r})=e^{i{3\over2}\theta}
\psi(\mathbf{r}e^{i\theta}),
\end{equation}
where the factor $e^{i{3\over2}\theta}$ comes from the three-dimensional volume
element \cite{Rein82}.
Under ${\hat U}(\theta)$, the Hamiltonian transforms as
\begin{equation}\label{complexr}
{\hat h}_\theta(\mathbf{r})={\hat U}(\theta){\hat h}
(\mathbf{r}){\hat U}(\theta)^{-1}.
\end{equation}
(This variant of  CS is referred to as {\it uniform complex scaling}; it
is to be distinguished from the {\it exterior complex scaling} discussed below.)
The transformed Hamiltonian 
${\hat h}_{\theta}$ is no longer hermitian as it acquires 
a complex potential.
However, for a wide class of  local and nonlocal potentials,
called {\em dilation-analytic potentials},
the so-called ABC theorem  \cite{Agu71,Bal71,Sim72})
is valid. This theorem states that:
\renewcommand{\labelenumi}{\roman{enumi}.}
\begin{enumerate}
\item
The bound states of  ${\hat h}$ and   ${\hat h}_{\theta}$ are
the same;
\item  
The positive-energy spectrum of the original Hamiltonian ${\hat
h}$ is rotated down
by an angle of $2 \theta$ into the complex-energy plane, exposing
a higher Riemann sheet of the resolvent; 
\item
The resonant states  of  ${\hat h}$ with eigenvalues $\tilde{E}_n$ satisfying
the condition $|arg(\tilde{E}_n)|<2\theta$  are also
eigenvalues of ${\hat h}_{\theta}$ and their wave functions are 
square integrable (see Fig.~\ref{cspoles}).
\end{enumerate}
%

Unfortunately, some of the phenomenological potentials  widely used
in nuclear physics are not dilatation-analytic (e.g., the Coulomb potential
of a uniformly charged sphere)  or they are  dilatation-analytic only
in a limited range of  $\theta$. For instance, the commonly
used  Woods-Saxon potential and its
derivative are dilatation-analytic only for 
$\theta < \theta_c=arctg({{a\pi}\over {R}})$.
Some of these difficulties, however, can be cured by the use of the {\em
exterior complex scaling} (ECS), originally  introduced in
Ref.~\cite{Gya71} to prove the existence of the norm of a Gamow
resonance for charged particles (see also Ref.~\cite{Sim79}). 
Within ECS, the rotation of $r$ to the
complex starts at a finite distance $r_a$ at or beyond the range of the
nuclear potentials $r_a\ge R$. The operator ${\hat U}_a(\theta)$ of the
exterior complex scaling is defined for a wave function $\psi(r)$ by
\begin{equation}
\label{extcs}
{\hat U}_a(\theta)\psi(r)=\cases{\psi(r)&, if  $ r \le r_a$\cr 
\psi(r_a + |r-r_a|e^{i\theta})&, if
$|r|>r_a$\cr}.
\end{equation} 
The ECS  influences only the asymptotic
behavior of functions and can be applied to any finite range
potential, including the  Coulomb potential in the region where it behaves
as $r^{-1}$.

The Berggren ensemble used in GSM contains  non square-integrable
functions; hence, a  regularization procedure 
is needed to calculate norms and
matrix elements needed in the GSM. The ECS can be used for this purpose.
The complex-scaled states have square
integrable wave functions and all related integrals are finite.
Since the resulting matrix elements are independent of the
angle, they
can be calculated at any convenient value of $\theta$.
However, if one is interested in the radial dependence of the wave
function for real $r$ (e.g., to estimate
partial widths), the CS transformation needs to be inverted,
\begin{equation}
\psi  ={\hat U}(\theta)^{-1}\psi_\theta.
\end{equation}
Such a back rotation
can  introduce large errors 
\cite{gyarbor}  if the solution
is expanded and approximated by a finite number of terms.
If Pad\'e approximants are used for 
performing back rotation \cite{Lef92},
good accuracy can be achieved for narrow resonances.
A comparison of the diagonalization in a Berggren basis  and the CS
method resulted in excellent agreement for the complex energy of the IAR in
the Lane model \cite{Bet08}. 

The CS and ECS methods have been widely used in many-body
calculations of unbound states and in the description of atomic and nuclear
reactions, see, e.g.,
Refs.~\cite{Kru88,Kru97,Kru01,Nic90,Res99,Mez07,LIs05,YHo79,Kru07}. 
For recent nuclear examples, see also Sec.~\ref{perspectives}.

\section{One-body space of the GSM}\label{onebody}

This Section describes  the s.p.~space of the GSM, i.e.,
the resonant and complex-$k$ scattering states generated by an auxiliary
s.p.~Hamiltonian.
 We describe the normalization procedure for
the continuum states and discuss the discretization of the scattering
contour. The optimal basis-generating potential is obtained from
the Hartree-Fock procedure generalized to resonant states. Finally,
selected examples of expansions in the Berggren basis are given. They
nicely demonstrate the completeness of the Berggren ensemble and the
accuracy of calculations.

\subsection{Resonant  and  scattering states of generating potential}

The radial s.p.~wave functions $u(k,r)$ that enter
the Berggren ensemble (\ref{Berggren_comp_rel})
are solutions of
the radial Schr{\"o}dinger equation:
\begin{eqnarray}
\frac{d^2 u(k,r)}{d r^2} = \left( \frac{\ell(\ell+1)}{r^2} + \frac{2 m}{\hbar^2} V(r) -
				     k^2 \right) u(k,r) \label{WS_Schr_eq},
\end{eqnarray}
where $\ell$ is the orbital angular momentum, 
$m$ is the reduced mass of the nucleon, 
and $V(r)$ is the  one-body  potential that generates the basis.
The choice of the potential is in principle arbitrary, but in practice, one is trying
to optimize it by adjusting its parameters to experimental s.p.~states. 
Another option to calculate it is using the Hartree-Fock (HF) method, as discussed
in Sec.~\ref{GHF_section}.

In many applications, a Woods-Saxon (WS) central field is used, supplemented by 
the  spin-orbit and Coulomb potentials:
\begin{equation}
V(r) = -V_0 f(r) - 4 V_{so} (\mathbf{l \cdot s}) \frac{1}{r} 
       \left| \frac{df(r)}{dr} \right| + V_c(r),
\label{WS_pot}
\end{equation}
where
\begin{equation}
\label{vagottWS}
f(r)=\left\{
\begin{array}{rl}\left[ 1 + \exp \left( \frac{r-R_0}{d} \right) \right]^{-1}
&\textrm{, if } r~<~R\\
0&\textrm{, if } r~\geq~R ~,
\end{array}
\right.
\end{equation}
is a WS form-factor, 
$V_0$ is the WS potential  strength,
$R_0$ is the  radius, $d$ is the WS potential diffuseness, $R$ is a  cutoff radius,
$V_{so}$ is  the spin-orbit coupling strength, 
and $V_c(r)$ is the Coulomb potential for protons.
The solution $u(k,r)$ is regular at the origin, i.e., 
\begin{equation}
u(k,r) \sim C_0 r^{\ell+1} \mbox{ , } r \rightarrow 0.
\end{equation}
At large distances, where the nuclear part of the potential is zero,
 $u(k,r)$  satisfies  the asymptotic radial equation:
\begin{eqnarray}
\frac{d^2 u(k,r)}{d r^2} = \left( \frac{\ell(\ell+1)}{r^2} + {{2\eta k}\over r} -
				     k^2 \right) u(k,r) \label{assimpt_eq},
\end{eqnarray}
whose  solutions are
the regular and irregular Coulomb functions $F_l(kr,\eta)$ and $G_l(kr,\eta)$,
respectively, and $\displaystyle \eta = \frac{m Z}{k \hbar^2}$ is 
the Sommerfeld parameter.
The outgoing and incoming Coulomb waves are usually denoted as
$H^{\pm}_{\ell \eta}=G_l(kr,\eta)\pm iF_l(kr,\eta)$.
 For neutrons, $\eta =0$ and the asymptotic solutions are the  Hankel functions. 
 The  boundary condition at $r\ge R$ can be written as
\begin{eqnarray}
u(k,r) &\sim& C_+ H^+_{\ell \eta} (kr) \mbox{ , } r\ge R \, \mbox{(resonant states)},
 \\
u(k,r) &\sim& C_+ H^+_{\ell \eta} (kr) + C_- H^-_{\ell \eta}
 (kr) \mbox{ , } r\ge R\,  \mbox{(scattering states)},\label{boundary_cond}
\end{eqnarray}
where $C_0, C_+$, and $C_-$ are normalization constants. 
For the resonant states, i.e., the $S$-matrix poles, the wave functions have a purely 
outgoing character;  they form a discrete component of the Berggren ensemble.

In most applications, Gamow states are expressed in the  coordinate
representation; they are found by 
direct integration of the Schr{\"o}dinger equation with 
outgoing boundary conditions. 
This procedure is straightforward for spherical local
potentials
and can be applied to deformed potentials by using  a coupled-channel
decomposition of the wave function \cite{Kru00}. 
In some cases, however, 
it is useful to work with Gamow states in momentum representation  \cite{Her84,Mon91}. 
The continuum discretization in $k$-space can be 
carried out using the Fourier-Bessel transform. 
The Schr{\"o}dinger equation in momentum representation reads:
\begin{equation}\label{Schrodinger_eq_momentum}
\frac{\hbar^2 k^2}{2m} u(k) + \int_{0}^{+\infty} V(k,k') u(k')~dk'
 = \tilde{E}u(k), 
\end{equation}
where $V(k,k')$ is the Fourier-Bessel transform of the generating  potential.
Contrary to the coordinate representation,
where non-local potentials have to be treated with the
equivalent potential method \cite{Vau57},
local and non-local potentials are treated on the same footing
in the momentum representation.
Moreover, discretization of the
Schr{\"o}dinger equation can be replaced  by 
matrix diagonalization \cite{Hag06a}.
Recent applications of the momentum representation can be found in
Refs.~\cite{Hag05,Hag06}. This way of generating of Gamow states 
turned out to be very effective, especially for 
well-deformed nuclei.

\subsection{Normalization of single-particle states}\label{norm1body}

The s.p.~states in (\ref{WS_Schr_eq}) need to be properly normalized,
either to unity (resonant states) or to  Dirac delta
(scattering states).

The normalization to the Dirac delta 
 is straightforward to achieve, as it is
equivalent to the condition   $2 \pi C_+ C_- = 1$ in Eq.\,(\ref{boundary_cond}),
which comes from  the asymptotic form of  Hankel/Coulomb wave functions. 
The values of the constants $C_+$ and $C_-$ can be obtained 
from the matching condition at
$r=R$: 
\begin{equation}\label{Cp_Cm_def}
\frac{d}{dr}\left[C_+ H^+_{\ell \eta} (kR) + C_- H^-_{\ell \eta}
(kR)\right] = \frac{d}{dr}u(k,R),
\end{equation}
where $u(k,R)$ is  determined by the direct integration from $r=0$ to $r=R$.

The normalization of decaying  states is, however, more complicated as,
contrary to bound states, they diverge in modulus for 
$r \rightarrow +\infty$. A solution to this problem can be obtained 
by considering the analytic property of the norm of $u(k,r)$, denoted as 
$\mathcal{N}(k)$. If $k$
is  positive imaginary, i.e.,  the state  $u(k,r)$ is bound, 
 $\mathcal{N}(k)$ is given by
\begin{eqnarray}
\mathcal{N}(k) = \sqrt{\int_0^{+\infty} u(k,r)^2~dr} \label{Nk_bound_state}.
\end{eqnarray}
For resonant states with a nonzero real part,
the exterior complex 
scaling described in Sec.~\ref{complexsc} can be used:
\begin{eqnarray} \label{Nk_complex_scaling}
\int_0^{+\infty} u(k,r)^2~dr &=& \int_0^{R} u(k,r)^2~dr + 
C_+^2 \int_{R}^{+\infty} H^+_{\ell \eta} (kr)^2~dr \nonumber \\ 
                             &=& \int_0^{R} u(k,r)^2~dr + 
 C_+^2 \int_{0}^{+\infty} H^+_{\ell \eta}
  (k R + kx e^{i \theta})^2~e^{i \theta}~dx,
\end{eqnarray}
where $\displaystyle 0 < \theta < \frac{\pi}{2}$ is the angle of complex
rotation.  Analyticity of $u(k,r)$ with respect to $r$ implies that
 the integral (\ref{Nk_complex_scaling}) is
  independent of $R$ and $\theta$. Indeed, if $u(k,r)$ is a
decaying  state, there always exists a minimal angle $\theta_c > 0$ for which
the complex integral of Eq.~(\ref{Nk_complex_scaling}) converges for
$\theta > \theta_c$, to a value independent of $R$ and $\theta$.
Hence, one can define the analytic continuation of $\mathcal{N}(k)$ as
the square root of Eq.~(\ref{Nk_complex_scaling}).

The normalized  s.p.~wave function 
$\psi(k,\mathbf{r})$ becomes:
\begin{equation}
\psi(k,\mathbf{r}) = \frac{u(k,r)}{r} \bigl [Y_{\ell}( \hat r )\chi_s \bigr]_{jm}.
\end{equation}
All the angular momentum/isospin algebra
of the GSM is carried out exactly in the same way as in the standard SM.
The  difficulties inherent to the Berggren basis
are all contained in the radial wave function  $u(k,r)$.

\subsection{Discretization of the scattering contour}\label{cdisc}

In practical applications, as 
the  s.p.~basis must be finite,
the contour integral along $L^+$ is
limited to the finite range $k < k_{max}$ and then discretized
according to a preferred quadrature.
The momentum cut-off $k_{max}$ must
be sufficiently large to provide convergence. 
The resulting  {\it discretized
completeness relation } reads:
\begin{equation}\label{Berggren_discr}
\sum_{n \in (b,d)} | u_n \rangle \langle u_n |  + 
\int_{L^+} |u(k) \rangle \langle u(k)| dk \simeq 
\sum_{n \in (b,d)} | u_n \rangle \langle u_n | +  \sum_{i} \omega_i
|u(k_i) \rangle \langle u(k_i)| = \sum_{i} |u_i \rangle \langle u_i|,
\end{equation}
where $(k_i,\omega_i)$ is the set of
discretized momenta and associated weights provided by a 
quadrature  (usually, a Gauss-Legendre quadrature is chosen).
The discretized  scattering
 states 
  \begin{equation}\label{dscat}
 u_i(r) = \sqrt{\omega_i} u(k_i,r)
 \end{equation}
are normalized to unity as the discretization implies  the replacement
of $\delta(k_i - k_{i'})$ by $\omega_i \delta_{i i'}$. The  relation
(\ref{Berggren_discr}) involves both resonant and discretized scattering
states; it is formally identical to the standard completeness relation
in a discrete basis.

\subsection{Gamow Hartree-Fock potential} \label{GHF_section}

The nature of s.p.~resonant states entering the
Berggren ensemble  changes with the depth  of the generating
potential. Obviously, a broad resonant state  becomes progressively
narrow and eventually bound as the potential's depth increases. This implies that,
in order to optimize the s.p.~basis,
the generating potential should  be chosen according to the  nucleus
studied. Let us consider an example of the He chain, treated in a $^4$He
core+valence particle framework, discussed later in this paper.
The potential simulating the $^4$He core is fitted to reproduce the
s.p.~resonances in $^5$He. While the resulting basis will
work well for  a two-neutron  halo nucleus $^6$He, it is not well suited
for a four-neutron halo  system  $^8$He.

An optimal generating potential can be obtained by means of the 
 HF method from the underlying many-body
Hamiltonian. However, in order to be used in the completeness relation 
(\ref{Berggren_comp_rel}), the HF potential 
 has to be spherical and the HF approach has  to be
extended to unbound systems. This approximation is referred to as the 
Gamow Hartree-Fock  (GHF) \cite{Mic04}. The starting point of GHF is a
GSM  Hamiltonian 
\begin{equation}\label{H_def}
H_{GSM} = \sum_i \hat{h}_i + \sum_{i < j} \hat{V}_{ij},
\end{equation}
consisting of  a one-body term 
\begin{equation}\label{onebodyp}
\hat{h} = \frac{\hat{p}^2}{2 m} + U(r)
\end{equation}
 and a residual two-body
interaction $\hat{V}_{ij}$ acting among valence particles.
The one-body field
 $U(r)$ in Eq.~(\ref{onebodyp}) represents
the core-valence interaction. It can be taken in a WS form 
(\ref{WS_pot}).
In order to impose spherical symmetry, a uniform-filling
approximation can be employed, in which nucleons 
occupy uniformly the valence shell
(i.e., no individual HF orbitals are blocked).
The matrix
elements of the HF potential $U_{uf}$
 between two spherical states  $\alpha$ and $\beta$  carrying quantum
numbers  $(\ell,j)$ are:
\begin{eqnarray}
\langle \alpha | \hat{U}_{uf} | \beta \rangle = \langle \alpha | \hat{h} | \beta \rangle
                                        + \frac{1}{2 j + 1} \sum_{m,\lambda,m_{\lambda}} \frac{N(\lambda)}{2 j_{\lambda} + 1} 
					 \langle \alpha m ~ \lambda m_{\lambda} | \hat{V} | \beta m ~ \lambda m_{\lambda} \rangle \label{U_uf},
\end{eqnarray}
where $\lambda$ is an occupied shell having quantum numbers
$(\ell_{\lambda},j_{\lambda})$, $N(\lambda)$ is the number of particles
occupying the $\lambda$-shell, and $m$, $m_{\lambda}$ run over all
possible values. One recalls that $U_{uf}$ provides the exact HF
potential only for closed-shell nuclei, with which all 1p-1h excitations
from the GHF ground state vanish.

If the nucleus has only one particle or hole outside closed
shells, it is possible to define a better approximation to $U_{uf}$,
which has been called $M$-potential in Ref.~\cite{Mic04} and is denoted
as $U_M$. In this case, one blocks the last particle or hole in the
one-body state of largest angular projection $m$, so that the HF ground
state is coupled to $J$=$m$.
The spherical $M$-potential is defined by
averaging the  HF potential over the magnetic quantum number $m$:
\begin{eqnarray}
\langle \alpha | \hat{U}_M | \beta \rangle = \langle \alpha | \hat{h} | \beta \rangle
                                        + \frac{1}{N_{\ell j}} \sum_{m = j+1-N_{\ell j}}^{j} 
					  \sum_{\lambda,m_{\lambda}} 
					  \langle \alpha m ~ \lambda m_{\lambda} | \hat{V} | \beta m ~ \lambda m_{\lambda} \rangle \label{U_M},
\end{eqnarray}
where $N_{\ell j}$ is the number of nucleons occupying the valence shell
$(\ell, j)$.  
It is straightforward to show that all
1p-1h excitations from the GHF ground state involving the  particle in
the unfilled shell vanish in this case. While this does not hold 
for the hole states,  one can still expect the $M$-potential to perform 
better than $U_{uf}$ in this case.

If the HF ground state is unbound, one is forced to disregard the
imaginary part of the  GHF potential. For the case of one valence particle,
the imaginary component of the GHF potential disappears due to
antisymmetrization, and the  imaginary part generated by the core shells
must  be spurious, as the core is assumed to be well bound. 
Even though a removal of an imaginary part of the HF potential slightly
affects  self-consistency,  solutions for unbound nuclei
can easily be obtained. Other aspects of the
GHF procedure remain unchanged as compared to the standard HF procedure
for bound systems.

Within the mean-field theory, pairing correlations can be included by means of the 
Hartree-Fock-Bogoliubov method (HFB). Its resonant-state extension is called
Gamow-HFB \cite{Mic09}.
For a bound nucleus,  upper and lower components of one-quasi-particle
Gamow states  meet   outgoing
boundary conditions \cite{Mic09,Bel87}.
While the  Gamow-HFB approach allows for the study of the structure of 
resonant states in weakly bound paired systems, its extension
to unbound nuclei still needs to be worked out, even
though an approximate scheme has been devised in Ref.~\cite{Mic05a}.

Another complex-energy, self-consistent approach that can be used to
study  particle unstable nuclei is the
complex-scaled  HF (CSHF) method \cite{Kru97}. In this
approach, the HF potential is complex-rotated (\ref{complexr})
in order to have all its occupied states square-integrable.
Contrary to GHF, self-consistency is fully maintained for unbound
nuclei, because the CSHF potential is complex. 
However, one cannot extend CSHF beyond the HF level. For that,
scattering states would have to be calculated in all
directions of the complex energy plane in order to extend CS to
non-resonant continuum. This is impossible because the CSHF potential
diverges exponentially in some regions of the complex plane. Consequently, 
as CSHF cannot be used to generate a GSM basis, and pairing correlations cannot be
treated within CS-HFB,  the range of applications of CSHF is rather limited.

\subsection{Numerical tests of the Berggren completeness relation}\label{one_body_calc}

In this Section, we illustrate the validity and accuracy of
Eq.~(\ref{Berggren_comp_rel}) by means of several
numerical examples. To this end,
we consider the expansion of 
resonant states generated by a WS potential $V_E(r)$ 
(\ref{WS_pot})
with a
Berggren basis generated by a WS potential $V_B(r)$ of different  
 $V_0$ strength, but
identical for all remaining  parameters. 
The s.p.~Hamiltonian to be diagonalized $\hat{h}_E$
is
represented by a continuous matrix in the eigenstates of
$\hat{h}_B=\displaystyle \frac{\hat{p}^2}{2m} + V_B$:
\begin{eqnarray}
\langle u_n | \frac{\hat{p}^2}{2m} + V_E
 | u_p \rangle 
&=& \frac{\hbar^2 k_n^2}{2m} \delta_{n p} + \int_{0}^{+\infty} u_n(r) 
\Delta(r) u_p(r)~dr \nonumber, \\
\langle u(k) | \frac{\hat{p}^2}{2m} + V_E | u(k') \rangle 
&=& \frac{\hbar^2 k^2}{2m} \delta(k-k') + \int_{0}^{+\infty} u(k,r) \Delta(r) u(k',r)~dr \nonumber, \\
\langle u_n | \frac{\hat{p}^2}{2m} +  V_E | u(k) \rangle &=& \int_{0}^{+\infty} u_n(r) \Delta(r) u(k,r)~dr \label{cont_matrix},
\end{eqnarray}
where $u_i(r)$  and  $u(k,r)$ are, respectively, 
 resonant  and scattering states 
of the Berggren basis generated by $\hat{h}_B$,
and $\Delta(r) \equiv V_E(r) - V_B(r)$. 

Following the contour discretization described in Sec.~\ref{cdisc},
the continuous representation of (\ref{cont_matrix}) is 
approximated by a  complex symmetric matrix:
\begin{eqnarray}\label{hspm}
\langle u_{i} | \frac{\hat{p}^2}{2m} + V_E | u_{i'} \rangle 
= \frac{\hbar^2 k_i^2}{2m} \delta_{i i'} + \int_{0}^{+\infty} u_i(r) \Delta(r) u_{i'}(r)~dr \label{discr_matrix},
\end{eqnarray}
where $u_i(r)$  are discrete vectors
representing both resonant and scattering states. 
The eigenvectors of (\ref{hspm}) represent  the
resonant  and discretized scattering states  of $V_E$. 

As an illustrative  example, let us  consider the expansion of a neutron 
$2p_{3/2}$
eigenstate of a WS potential (\ref{WS_pot}) in a Berggren basis.
The radial wave function of the expanded  state, denoted as
$u_E(r)$, is a linear combination of $p_{3/2}$ basis states $u_B(k,r)$:
\begin{eqnarray}\label{ue_expansion}
u_E(r) = \sum_{n \in (b,d)} c_n u_B(k_n,r) + 
\int_{L^+} c(k) u_B(k,r) dk, 
\end{eqnarray}
where $c_n$ and $c(k)$ are the  expansion coefficients. As discussed
in Sec.~\ref{cdisc},
 $L^+$  has to be discretized.

%
\begin{figure}[htb]
\centerline{\includegraphics[trim=0cm 0cm 0cm 0cm,width=0.4\textwidth,clip]{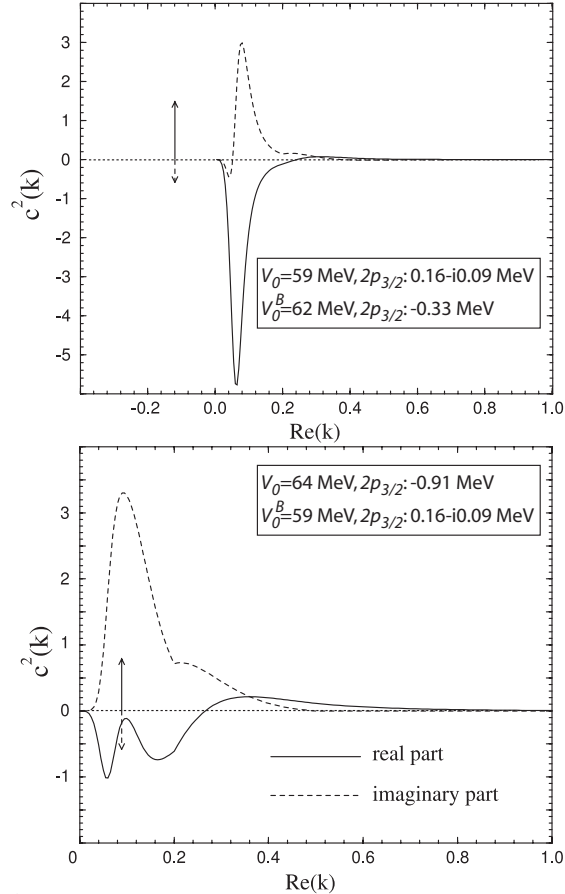}}
\caption{
Distribution of the squared amplitudes $c^2(k)$
of the neutron state $2p_{3/2}$ of the WS  
in the basis generated by another  WS
potential.  The amplitudes of both real (solid line)
and imaginary (dashed line) parts of the wave function
are plotted as a function
of Re[$k$]. The height of the arrow gives the squared amplitude
of the $2p_{3/2}$ resonant state  contained in the basis.
The complex contour $L^+$ corresponds to three straight
segments in the complex $k$-plane, joining the points:
$k_0$=0.0-i0.0, $k_1$=0.2-i0.2, $k_2$=0.5-i0.0,  and $k_3$=2.0-i0.0,
all in units of fm$^{-1}$. 
Top: $V_0$=59 MeV, $V_0^{(B)}$=62 MeV. Here the expanded state
is a $2p_{3/2}$ resonance. Bottom: 
 $V_0$=64 MeV, $V_0^{(B)}$=59 MeV. Here the expanded state
is bound. (Example taken from Ref.~\cite{Mic03} where 
details of the calculation can be found.)
}
\label{example1}
\end{figure}
The expansion coefficients (\ref{ue_expansion}) are shown 
in Fig.~\ref{example1}
where the depth of the WS potential
is varied to generate a   resonant or bound  $2p_{3/2}$ state.
In both cases, 
the importance of the
scattering continuum is noticeable. This example  nicely
demonstrates the invalidity of the pole approximation to provide a
quantitative description of weakly bound or resonant states.

More examples of s.p.~expansions can be found in Refs.~\cite{Mic03} 
(neutrons) and  \cite{Mic04} (protons). Let us only mention that in the proton
case the role of the scattering continuum is somehow reduced due
to the confining effect of the Coulomb barrier, especially at small
energies. Still, for  a precise description of
eigenstates, the scattering  component needs to be included.

\section{Gamow Shell Model}\label{GSMS}

\subsection{Many-body GSM basis}

The discretized basis (\ref{Berggren_discr}) can be used to
extend  the completeness relation to  the many-body
case, in a full analogy with the standard SM in
a complete discrete basis, e.g., the HO
basis. The many-body basis of  GSM  corresponds to 
Slater determinants (SD) built from one-body
states constituting the  Berggren ensemble:
\begin{equation}\label{Berggren_SD}
| SD \rangle_n = |u_{i_1} \cdots u_{i_A} \rangle,
\end{equation}
where $|u_i \rangle$ are  properly normalized one-body states
(either resonant or scattering) of a generating potential.
The completeness of the Berggren ensemble guarantees the 
closure relation for the many-body GSM
basis:
\begin{equation}\label{mbody}
\sum_{n} | SD_{n} \rangle \langle SD_{n} | \sim 1 \label{many_body_Berggren_compl_rel},
\end{equation}
where $n$ runs over all possible SDs (\ref{Berggren_SD}). 
The approximate equality in (\ref{mbody}) is a consequence of
the continuum discretization.

\subsection{Determination of many-body resonant states} \label{determination}

The GSM  Hamiltonian matrix in the basis (\ref{mbody}) is
complex symmetric. In a standard  SM, the Lanczos method
is often used to determine the eigenstates of the
Hamiltonian matrix. It allows calculating selected states, e.g.,  ground state
and low-lying excited states, in very large configuration spaces.
However, the Lanczos method cannot be applied mutatis mutandis to the GSM matrix
problem, due to the presence of scattering states. As many-body resonant
states are embedded in  the discretized continuum of many-body scattering
states, the lowest-energy  principle cannot be used to guide our choice.

One way of selecting many-body resonant states was
proposed in Refs.~\cite{Bet02,Bet03} dealing with systems
having two valence nucleons. Here, the selection
of  two-particle resonances is done with  a suitable choice of the
 contour $L^+$.  It was realized that by choosing a square-shaped contour, 
the resonant states appear in the so-called allowed
region of the complex-energy plane \cite{Bet02} and this facilitates their identification.

\begin{figure}[htb]
\centerline{\includegraphics[trim=0cm 0cm 0cm 0cm,width=0.5\textwidth,clip]{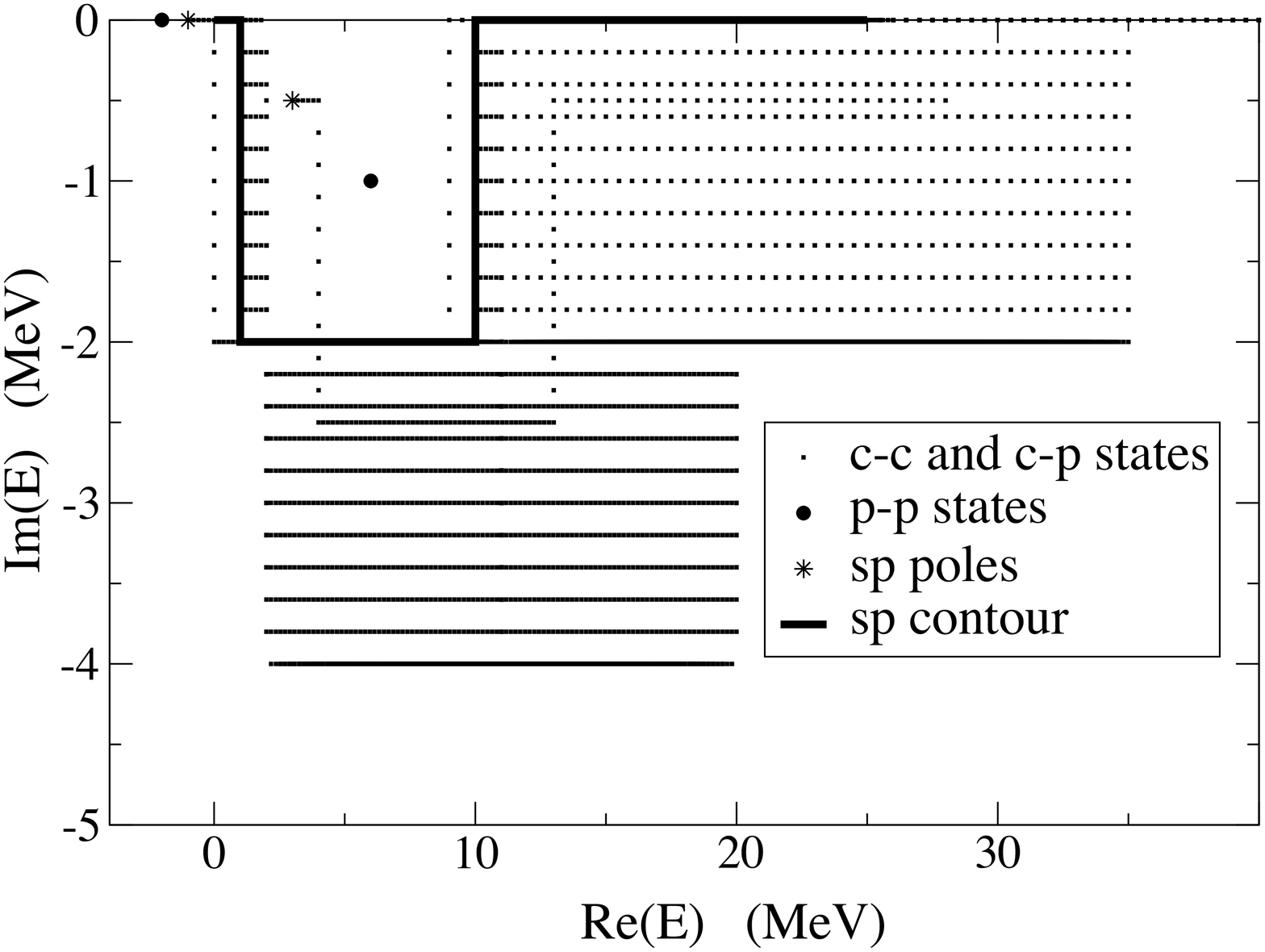}}
\caption{
A typical square-shaped contour in the complex energy-plane
used for calculation of  two particle resonant states in Ref.~\cite{Bet02}.
Stars mark the s.p.~pole (p) states, while the
thick line denotes the scattering 
contour (c). Full dots  denote the two-particle energies
of the pole-pole type and those in which at least one single particle
is in a scattering  state. (See Ref.~\cite{Bet02} for details.)}
\label{uncorr}
\end{figure}

In Fig.~\ref{uncorr}  the allowed region is a square-shaped area defined by a 
s.p.~contour. Due to the low density of  two-particle
scattering  states in this region,  the trajectories of  two-particle
resonances can be easily followed as the strength of the residual interaction
increases.
%
It has been observed that the two-body resonant states are mostly based
on the pole-pole (p-p) configuration, although the role of  discretized
continuum states (p-c, c-c) in Fig.~\ref{uncorr} is not negligible. The
result that the correlated two-particle resonant states are built mainly
from configurations formed from s.p.~resonant  states gives 
strong support for the use of the overlap method discussed below.
Unfortunately this convenient method might not be generalized easily to
the case of more than two valence particles.

Another  solution to the problem of resonant-state selection, particularly useful
in the case of several valence nucleons, has been
proposed in Ref.~\cite{Mic03}. It is based on 
the overlap method, utilizing the property  that the main
components  of many-body resonant states are built from  resonant
SD. In the first step, the GSM Hamiltonian is  diagonalized in a
smaller basis consisting of s.p.~resonant states only (a pole approximation).
Here, some variant of the Lanczos method can be
applied. The diagonalization yields the first-order
approximation to many-body resonances
$| \Psi_0 \rangle$.  In the second step, 
$| \Psi_0 \rangle$ is used  as a Lanczos pivot.
Diagonalization of the  GSM Hamiltonian in this subspace  yields
many-body eigenstates. The resonant state
is the eigenstate $|\Psi \rangle$ which maximizes the overlap 
$|\langle \Psi_0 | \Psi \rangle|$.
This method identifies  the  many-body resonant
state provided its coupling
to the non-resonant continuum  is weak, i.e.~10-30\%. This condition
is usually satisfied, especially if the one-body basis has been
optimized by way of the GHF procedure.

%
\begin{figure}[htb]
\centerline{\includegraphics[trim=0cm 0cm 0cm 0cm,width=0.7\textwidth,clip]{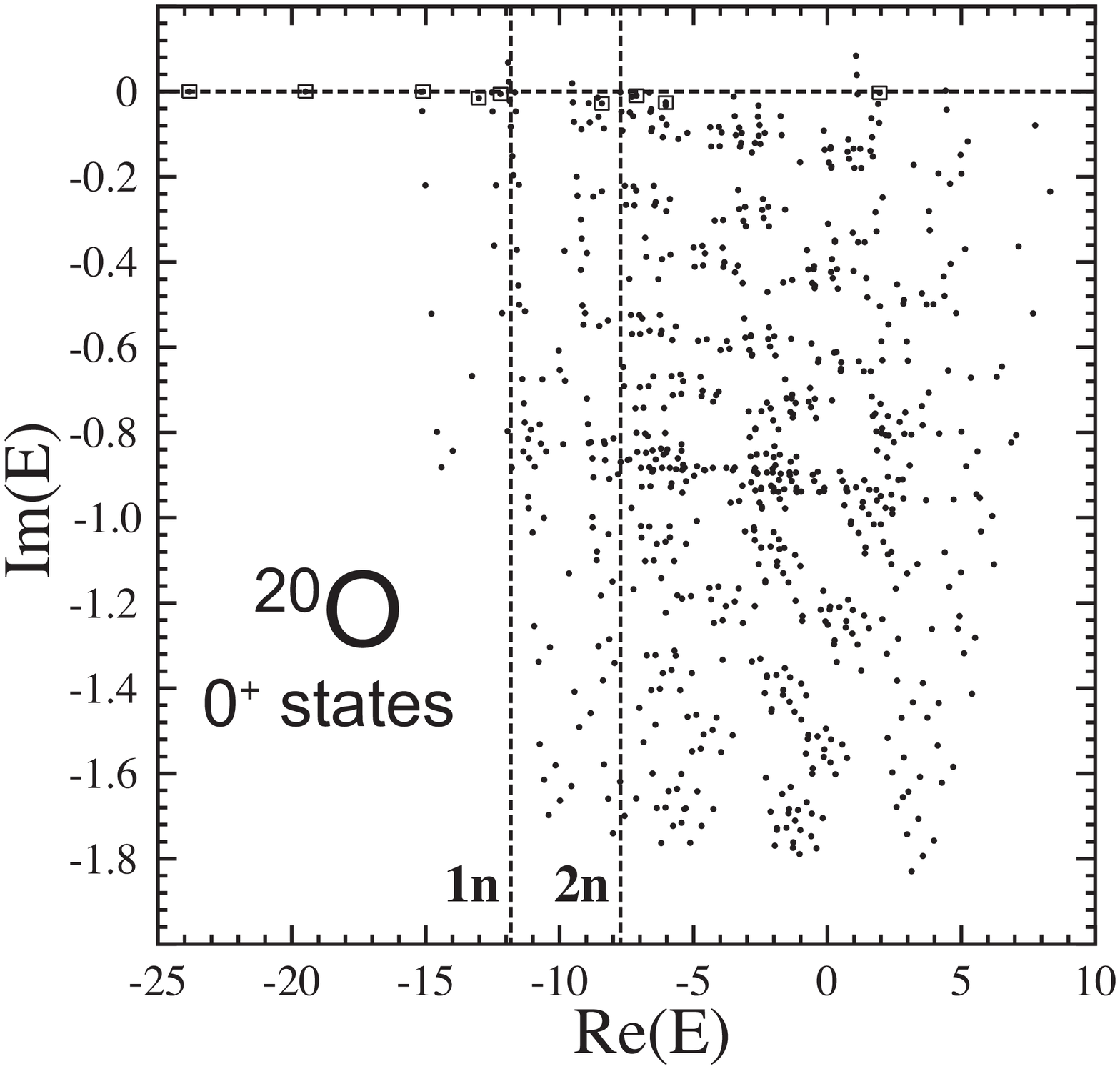}}
\caption{
Complex energies of the  $0^+$ states in $^{20}$O calculated in the GSM+SDI
model.
One- (1n) and two-neutron (2n) emission thresholds are indicated. 
The many-body resonant  states  (bound states and narrow
resonances) are
marked by squares. Note that all those state have zero or positive
widths, as expected. The remaining eigenstates represent the
non-resonant continuum. }
\label{overlap_method_O20}
\end{figure}
Fig.~\ref{overlap_method_O20} illustrates the overlap technique
for  $J^\pi = 0^+$ states in  $^{20}$O considered as a four-neutron system outside the
$^{16}$O core~\cite{Mic03}. Diagonalization of
the GSM Hamiltonian matrix provides  complex-energy eigenstates.
Compared with  Fig.~\ref{uncorr},
the number of many-body states is much larger and the 
regular pattern of non-resonant states reflecting the 
structure of the scattering contour is gone (the figure represents the projection
of four-dimensional trajectories onto two dimensional space). 
While the two lowest (bound) states can
be simply identified by inspection, for the higher-lying
states it is practically impossible to separate the resonances 
from the non-resonant continuum. However, the
procedure outlined above makes it possible to identify
unambiguously the many-body resonance states.
It is seen that the resonant states shown
in Fig.~\ref{overlap_method_O20} they all have negligible or very small positive
widths, as expected. 
 The many-body resonant states  appear to  be stable with 
respect to small changes  of the contour (as the physical 
solutions should not be dependent on the deformation of the basis).
On the other hand, the states representing the non-resonant many-body continuum
move in the complex energy plane with the contour's deformation.

\subsection{GSM Hamiltonian}

Hamiltonian matrix elements between resonant states can be
calculated with complex scaling (see Sec. \ref{complexsc}). 
However, this is not the
case with scattering states, for which complex-scaled integrals may not
converge regardless of  the angle of rotation, as is the case for the
 normalization of scattering states (see Sec. \ref{norm1body}).
In the latter situation, since the result is analytical (Dirac delta),
the problem can be dealt with by means of the  contour discretization
(see Sec. \ref{cdisc}).
However, it is very difficult to calculate matrix elements of general
interactions; hence, in the first applications, only localized
Hamiltonians were used. Examples are the Gaussian separable interaction
\cite{Bet02,Hag05}, the Surface Delta interaction  \cite{Mic02,Mic03}, and  
the Surface Gaussian interaction (SGI)
 \cite{Mic04} (see Sec.~\ref{H_scheme}). For these schematic forces, the
calculation of two-body  matrix elements is straightforward. 
In order to deal with realistic interactions, which are very complicated to
express and infinite-ranged in shell-model coordinates, a  new
approach, based on a HO basis expansion technique,  had to be
devised (see Sec. \ref{GSM_real_inter}).

\subsection{Matrix elements of electromagnetic operators}
 
In the GSM,  the  electromagnetic (EM) transition 
selection rules and the angular momentum and isospin algebra  do not change
compared to the standard SM. However, 
to calculate the EM transitions, one can no longer
use the long wavelength approximation because of the presence of
the non resonant continuum. Indeed, for the diagonal EM  matrix elements
$\langle n_i \ell_i j_i k_i|O| n_f \ell_f j_f k_f \rangle$ 
 ($k_i$ = $k_f$) {\it between the   scattering states}, 
 the complex scaling cannot be carried out.
 Furthermore, one cannot employ the long wavelength approximation to
the  EM  operators as they behave like $r^\lambda$.
Without the long wavelength approximation, however, the continuum-continuum
 matrix elements become finite, because it is always possible to carry out  a complex scaling
with the Bessel function of the photon $j_L(qr)$, as $q \neq 0$.
As all the other matrix elements can be regularized,
the EM  matrix elements are all well defined.

\subsection{Optimization of the scattering contour of GSM using the 
Density Matrix Renormalization Group method}

The application of  the Berggren ensemble to GSM  is associated with the
explosive growth in the number of configurations  with the number of
active particles and the size of the s.p.~space. To ensure completeness
of the Berggren basis, for each resonant 
s.p.~state $(\ell,j)$,  one should
include a large number of s.p.~scattering  states
$\{(\ell,j)^{(f)}; f=1,\dots,M\}_{\rm c}$  lying on a discretized
contour $L_+^{\ell_{j}}$.  All these  states
become new active shells; hence, the configuration space of GSM grows extremely
fast with the number of contour discretization points. 
Moreover, the resulting complex-symmetric GSM Hamiltonian
 matrix is significantly denser than the Hamiltonian
matrix of a conventional SM. Hence, taming the dimensionality growth in
solving the many-body Schr\"{o}dinger equation is crucial.

There are several specific features of the {\it dimensional catastrophe} in GSM
that may facilitate this task. Firstly, of all GSM eigenstates, of major
interest are the many-body resonant states. 
As discussed in Sec.~\ref{determination}, 
these eigenstates represent  a narrow class
of spatially localized states, immersed in a sea of non-resonant
scattering states.  Eigenstates of the GSM
Hamiltonian in the pole space can  serve as the reference vectors,
helping to identify resonant  eigenstates out of a huge space of all
GSM eigenvectors. Hence, instead of a  direct diagonalization of the
GSM Hamiltonian matrix,  a  strategy should be developed for an 
optimization of the scattering space.

The Density Matrix Renormalization Group (DMRG)
 method was introduced in Ref.~\cite{Whi92}
 to overcome the limitations
of Wilson-type renormalization groups to describe strongly correlated 1D
lattice  systems with short-range interactions  (see also
recent reviews  \cite{Duk04,Duk05}).   Up to now,
most of the DMRG studies were focused on equilibrium properties in
strongly correlated closed quantum systems (CQS) with a Hermitian density
matrix. Nevertheless,  non-equilibrium systems involving non-Hermitian
and non-symmetric density matrices can also be treated  in the DMRG
\cite{Car99}. The DMRG approach is ideally suited to optimize the size of the
scattering space in the GSM problem as the properties of the
non-resonant shells vary smoothly along the scattering contour. The main
idea of the GSM+DMRG truncation algorithm \cite{Rot06,Rot08} is
to gradually consider different s.p.~shells  of the discretized
non-resonant continuum in the configuration space and retain  only
$N_{\rm opt}$ optimal  states dictated by  the eigenvalues of the
density matrix with the largest modulus. 
Since the GSM Hamiltonian is complex symmetric, it is the
complex (or generalized)  variational principle \cite{moisey1,moisey,moisey2}
than the usual variational principle that 
governs the GSM+DMRG algorithm \cite{Rot08}.

The iterative GSM+DMRG procedure of constructing the eigenvectors is
divided into two phases: the warm-up phase and the sweeping phase. There
are two essential conditions which guarantee that the GSM+DMRG method
yields correct results. The first one is to ensure that all possible
couplings in the many-body wave function, allowed by the symmetries of
the problem, are present  in the warm-up phase. This means that one
should keep all the essential partial waves which are present in DMRG wave
functions, even though they may be absent at the pole approximation
level.
The second condition is
to ensure that the reference state $|\Psi_J\rangle^{(0)}$, which is a
key ingredient in selecting the target state $|\Psi_J\rangle$ from the
set of all GSM solutions of a given angular momentum and parity at each
iteration step, is correctly described.

The DMRG configuration space can be initially divided  into
 two subspaces: $A$ (the
reference subspace built from all s.p.~resonant shells 
$\{(n_1,\ell_1,j_1)$, $(n_2,\ell_2,j_2), \dots\}$ and one representative
non-resonant shell for each remaining  partial wave $\{(\ell_p,j_p),
(\ell_r,j_r), \dots\}$ in the valence space) that is essential to
produce many-body couplings, and $B$ (the
complementary subspace built from the s.p.~non-resonant shells
$\{(\ell_1,j_1)\}_{\rm c}$, $\{(\ell_2,j_2)\}_{\rm c}, \dots$). One
begins in the warm-up phase by constructing all product states
$|k\rangle_A$  forming the reference subspace $A$.

The many-body configurations  in $A$ can be classified in different
families  $\{n;j_A \}$ according to their number of nucleons $n$ and
total angular momentum $j_A$. 
All possible matrix elements of sub-operators of the two-body Hamiltonian
 acting in $A$ are calculated and stored. The Hamiltonian is then
diagonalized in $A$ to provide the  zeroth-order  approximation to the
targeted eigenstate $|J^{\pi}\rangle$, called the reference state
$|\Psi_J\rangle^{(0)}$. In the next step,  the subspace of the first
scattering shell $(\ell,j)_1$  belonging to the discretized contour $L^{+}$ 
is added. Within this shell, one constructs all possible many-body
states, denoted as $|i\rangle_B$ and grouped in  {\bf {$\{n_B;j_B\}$}}
families, and all matrix elements of sub-operators of the Hamiltonian.
This ensemble serves as a basis in which the GSM Hamiltonian is
diagonalized. The target state $|\Psi_J\rangle$ is  selected among the
eigenstates of the Hamiltonian as the one having the largest overlap
with the reference vector $|\Psi_J\rangle^{(0)}$. Based on  the
expansion:
\begin{equation}
|\Psi_J\rangle = \sum_{k_A,i_B} c^{k_A(j_A)}_{i_B(j_B)} \{ |k_A(j_A)\rangle \otimes  |i_B(j_B)\rangle \}^J~~,
\end{equation}
by summing over the  reference subspace $A$ for a fixed value of $j_B$, one 
obtains the reduced density matrix \cite{McC02}:
\begin{eqnarray}\label{rdm}
\rho ^{B}_{i_Bi'_B}(n_B,j_B) \equiv \sum_{k_A}c^{k_A(j_A)}_{i_B(j_B)} c^{k_A(j_A)}_{i'_B(j_B)}
\end{eqnarray}
which is complex-symmetric in the metric defining the Berggren ensemble.
In the warm-up phase, the reference subspace becomes the `medium' for
the `system' part in the $B$ subspace.

Truncation in the system part is dictated by  the density matrix. The
reduced density matrix is diagonalized and one retains  at most
${N}_{\rm opt}$ eigenstates of $\hat\rho^B$:
\begin{equation}\label{eigenrho}
\hat{\rho}^B (n_B,j_B)|\alpha\rangle_{B} = w_\alpha |\alpha\rangle_{B}
\end{equation}
having the largest nonzero values of  $|w_\alpha|$. Then, one expresses
the eigenstates $|\alpha\rangle_{B}$ in terms of the vectors
$|i\rangle_B$ in $B$:
\begin{equation}
|\alpha\rangle_{B}=\sum_{i} d^{\alpha}_{i}|i\rangle_B
\end{equation}
and all matrix elements of the sub-operators in these optimized states,
\begin{eqnarray}
_{B}\langle \alpha|O|\beta\rangle_{B}
=\sum_{i,i'}d^{\alpha}_{i}d^{\beta}_{i'}~ _{B}\langle i|O|i'\rangle_{B}  
\label{renorm}
\end{eqnarray}
are recalculated and stored. In a similar way, the warm-up procedure
continues by adding to the system part the configurations containing
particles in the next scattering shells  until the last shell in $B$ is
reached. At this point, all s.p.~states have been considered,  the
warm-up phase ends,  and the sweeping phase begins (cf. Refs. ~\cite{Rot06,Rot08}
for details).  The sequence of sweep-up and sweep-down phases
continues until convergence for target eigenvalues is achieved.

It has been demonstrated \cite{Rot06,Rot08} that the fully converged GSM results, with
respect to both the number of sweeps and the number of shells in the
discretized continua, can be obtained using the GSM+DMRG algorithm. 
Several attractive features have been found.
Firstly, the rank  $d_{\rm H}^{\rm max}$ of the biggest reduced density
matrix to be diagonalized in GSM+DMRG is practically constant with
respect to the number of shells $N_{\rm sh}$.  Hence, the ratio ${\rm
d}_{\rm H}^{\rm max}/D$, where $D$ is the dimension of the GSM
Hamiltonian matrix, decreases rapidly \cite{Rot06}. Secondly, the converged complex
eigenenergies depend weakly on the number of vectors at each step of the
iteration procedure, and the uncertainty of calculated eigenenergies
weakly depends on  $N_{\rm sh}$. This means that the convergence
features of the GSM+DMRG procedure can be tested by varying $N_{\rm sh}$
and $N_{\rm opt}$. Once those parameters  are optimized in 
`small-scale' GSM+DMRG calculations, the  final calculations can be
performed in the large model space to obtain fully converged results.
Thirdly, the  GSM+DMRG energy averaged over one sweep exhibits excellent
exponential convergence with  $N_{\rm opt}$, which allows to deduce the
asymptotic value with good precision. 
These
encouraging features of the GSM+DMRG algorithm  open the possibility for
systematic and high-precision studies of complex,  weakly bound, or
unbound nuclei, which require large configuration spaces for their
correct description.

\section{GSM applications}\label{GSMapp}

\subsection{GSM applications with schematic Hamiltonians} \label{H_scheme}

\begin{figure}[htb]
\centerline{\includegraphics[trim=0cm 0cm 0cm 0cm,width=0.5\textwidth,clip]
{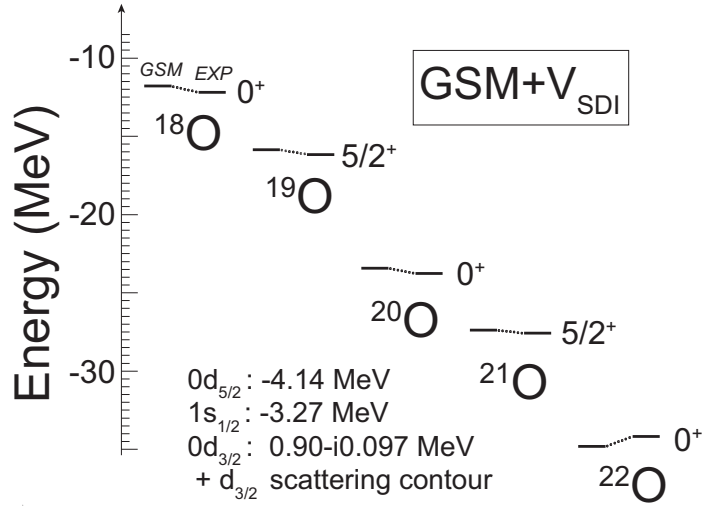}}
\caption{
Experimental (EXP) and predicted (GSM+$V_{SDI}$)  \cite{Mic03} binding energies of 
$^{18-22}$O. The energies are given with respect to the $^{16}$O core.
}
\label{O_SDI_chain}
\end{figure}

In the first applications of GSM,  
schematic core+valence Hamiltonians were applied.
In Ref.~\cite{Mic03}, neutron-rich isotopes of He and O have 
been studied  by approximating the residual interaction in
(\ref{H_def})  with the Surface Delta Interaction (SDI):
\begin{eqnarray}
V(\mathbf{r}_i,\mathbf{r}_j) = -V_{SDI} \delta(r_i - R_0) 
\delta(r_j - R_0), \label{SDI_def}
\end{eqnarray}
where $r_i = |\mathbf{r}_i|$, 
$R_0$ is the radius of the WS core potential, and $V_{SDI}$ is the SDI strength.
The main advantage of the zero-range SDI interaction is that
its  two-body matrix elements can be easily  calculated and no complex
scaling is involved. In the He case, the Berggren basis consisted of 
$0p_{3/2}$ and $0p_{1/2}$ resonant states  and the   associated scattering continua
$p_{3/2}$ and $p_{1/2}$. For the O isotopes, valence space
consisted of the bound $0d_{5/2}$,
$1s_{1/2}$ states, the  $0d_{3/2}$ resonance,  and the $d_{3/2}$
scattering contour. Since other scattering channels 
can only impact real energies, they
were  not included as  they could  effectively be taken into
account by renormalizing the SDI interaction strength.
Results for the chain of oxygen isotopes are shown in Fig.~\ref{O_SDI_chain}.

\begin{figure}[htb]
\centerline{\includegraphics[trim=0cm 0cm 0cm 0cm,width=0.7\textwidth,clip]{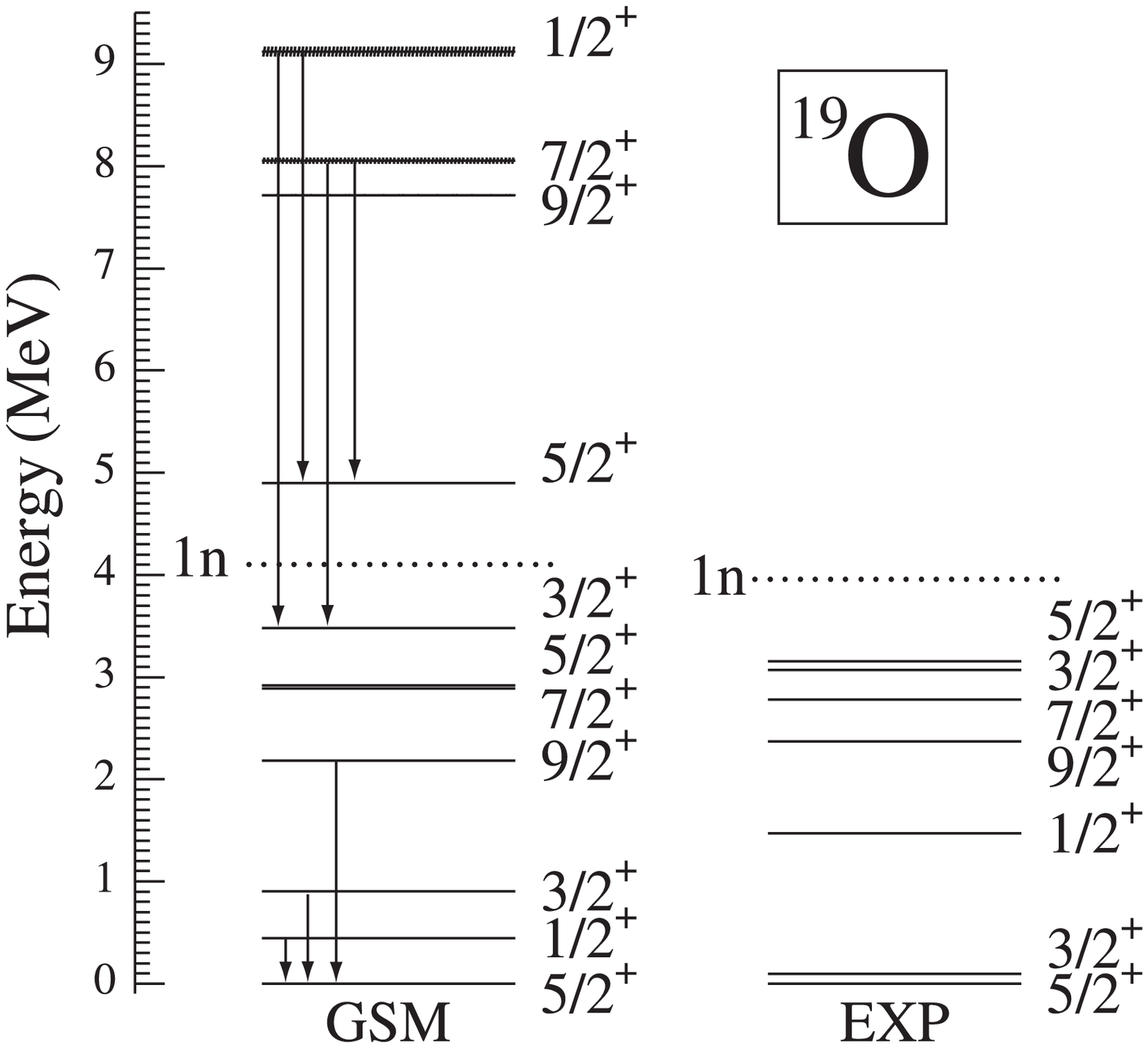}}
\caption{The GSM  level scheme of $^{19}$O 
calculated in the full $sd$ space of Gamow states and employing
the discretized (10 points) $d_{3/2}$ non-resonant continuum. 
The dashed lines indicate 
experimental and calculated one-neutron emission thresholds.
As the number of states becomes large above the one-neutron emission threshold,
  only selected resonances are shown.
The electromagnetic transitions discussed in \cite{Mic03}
are indicated by arrows. (See Ref.~\cite{Mic03} for details.)}
\label{O19}
\end{figure}

An early example of spectroscopic studies  with GSM
is shown in Fig.~\ref{O19} which displays 
the  level scheme  of $^{19}$O calculated with GSM+SDI.
The main 
experimental features are reproduced. In addition to energy levels, 
calculations were carried out for electromagnetic $E2$ and $M1$
transition probabilities.
%

\begin{figure}[htb]
\centerline{\includegraphics[trim=0cm 0cm 0cm 0cm,width=0.7\textwidth,clip]
{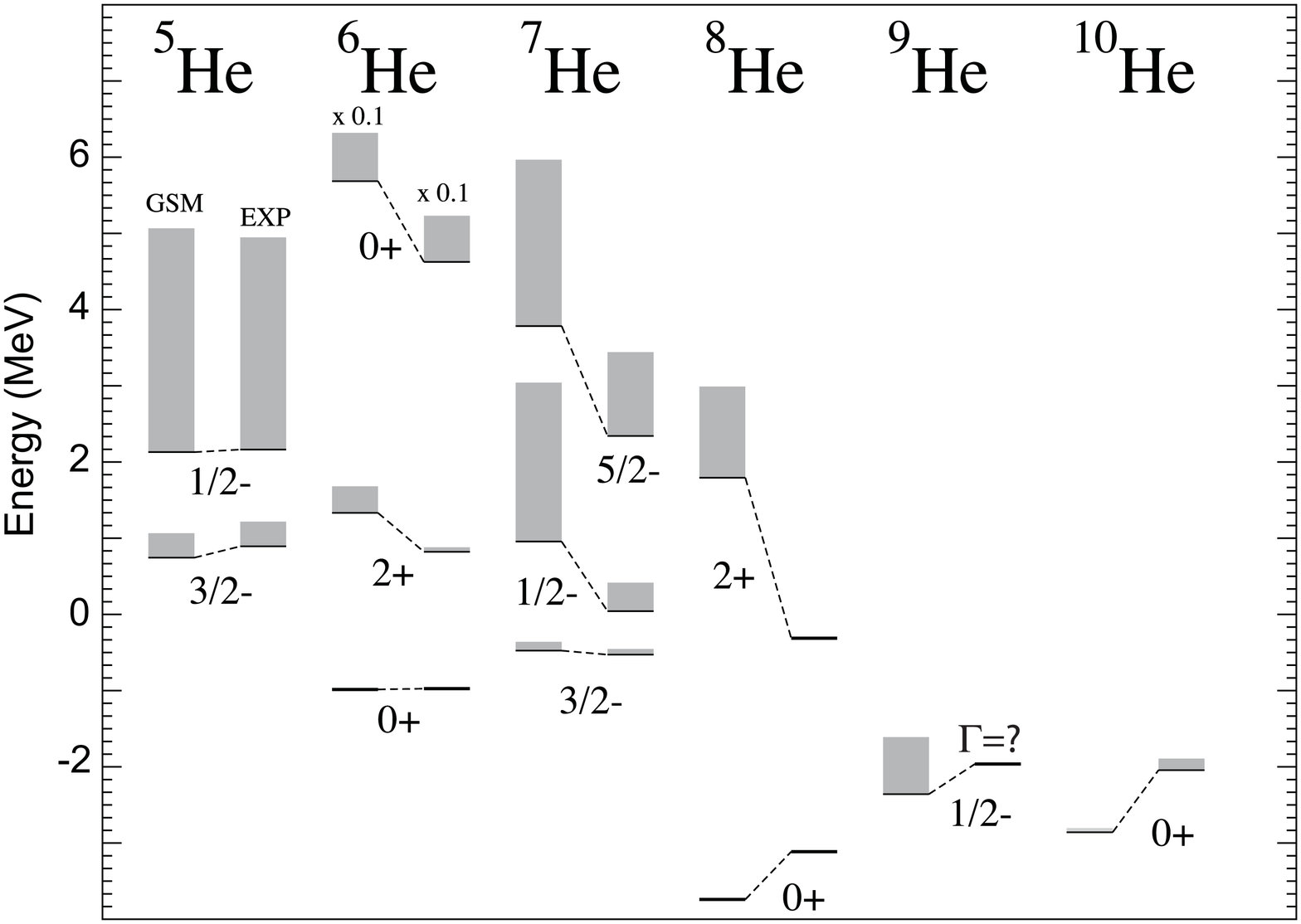}}
\caption{
Experimental (EXP) and predicted (GSM)  binding energies and 
spectra of helium isotopes obtained with the SGI Hamiltonian.
The resonance widths are indicated by shading.
The energies are given with respect to the core of $^{4}$He.
The ground-state width of $^9$He is not known experimentally; hence, it is
not indicated.
}
\label{He_SGI_chain}
\end{figure}

The GHF technique of basis optimization cannot be used effectively with SDI.
Hence, in order to improve  convergence of results, it was necessary to
employ another interaction. For that, a delta component of the SDI
interaction was replaced with  a Gaussian form factor, thereby
providing finite-range, while the second delta
function remains in order to keep the interaction surface-peaked.
The resulting  surface Gaussian interaction
(SGI) reads:
\begin{eqnarray}
V(\mathbf{r}_i,\mathbf{r}_j) = -V^{(J,T)}_{SGI} \exp 
\left[ -\left( \frac{\mathbf{r}_i - \mathbf{r}_j}{\mu} \right)^2 \right] 
                         \delta(r_i + r_j - 2 R_0), \label{SGI_def}
\end{eqnarray}
where $\mu$ is the range,  $V^{(J)}_{SGI}$ is the strength in the $JT$ 
channel, and all other values are defined similarly to
Eq.~(\ref{SDI_def}). Two-body matrix
elements can still be conveniently calculated with the SGI interaction, as
its radial components are one-dimensional finite integrals. The fact
that only one delta function appears in Eq.~(\ref{SGI_def}) implies that the 
GHF-SGI potential is well defined in coordinate space, 
albeit of non-local character. Non-locality of spherical
potentials is conveniently treated via the equivalent potential method
\cite{Mic04}, in which  a non-local potential is replaced by a local but
state-dependent potential.
The results of  GSM-SGI calculations for the neutron-rich He isotopes are
shown in Fig.~\ref{He_SGI_chain}. 
One obtains a reasonable description of
the ground-state(g.s.) energies and excited states relative to the g.s.~energy of
$^{4}$He.
%

A similar calculation for the lithium chain \cite{Mic04} suggests that
monopole corrections have to be added to the $T=0$ matrix elements of
SGI,  modeled through  a particle number dependence of 
$V^{(J,T=0)}_{SGI}$. While experimental data for $^6$Li and $^7$Li are
fairly reproduced, the drip-line nuclei $^{10}$Li and $^{11}$Li are
predicted to be overbound. This is due to a rather  limited
configuration space used in Ref.~\cite{Mic04}; most importantly, the
neglect of $s$-waves. Clearly, for a qualitative description of the
Borromean halo nucleus  $^{11}$Li, an extended valence space and more
realistic interactions are needed, as well as a  proper treatment of the
center-of-mass motion.

\subsection{GSM with realistic interactions} \label{GSM_real_inter}

The inclusion of realistic interactions in GSM has been a theoretical
and practical challenge. 
Standard SM methods rely on the $G$-matrix formalism
\cite{Hjo95} and Lee-Suzuki  similarity transformation
\cite{Suz80}. The Lee-Suzuki projection operator formalism or
the renormalization group technique can be used to 
derive low-momentum interactions, usually denoted as
$V_{low-k}$ \cite{Bog03}. The extension of the  
Lee-Suzuki formalism to the complex-energy GSM framework  has been done in 
Ref.~\cite{Hag05} where applications to Gaussian-type interactions
can be found.

Contrary to  schematic interactions  used in early GSM calculations, 
realistic interactions have a very complicated
structure \cite{Wir95,Mac01,Epe06}. They are defined in a
center-of-mass frame, so that the transformation 
to the  laboratory frame has to be carried out. 
For a HO basis, this  can be done by using  the 
Brody-Moshinsky coefficients  \cite{Law80}.
However, the generalization of this scheme to arbitrary bases through
the
vector-bracket transformation
\cite{Bal69,Won72,Kun79} is very time
consuming, as the finite  sums  in Moshinsky transformation must be
replaced by sums of two-dimensional momentum integrals. 
For the multipole decomposition  of the realistic interaction,
radial two-dimensional integrals have to be calculated with complex
scaling. Unfortunately, for some integrals, CS is impossible to apply. 

The solution to all these problems has been proposed in  Ref.~\cite{Hag06a}.
It is based on the idea that the nuclear interaction $\hat{V}$  is 
localized in space
around the nucleus, so that  the HO expansion should 
quickly converge for bound states
and narrow resonances. Expanded in a HO basis, two-body matrix
elements of $\hat{V}$ can be written as:
\begin{eqnarray}\label{Vosc}
\langle a b | \hat{V} | c d \rangle = \sum_{\alpha \beta \gamma \delta}^{N_{\rm max}} 
                                \langle \alpha \beta | \hat{V} | \gamma \delta \rangle 
				\langle a | \alpha \rangle    \langle b | \beta \rangle     
				\langle c | \gamma \rangle    \langle d | \delta \rangle,  \label{V_HO_exp}
\end{eqnarray}
where $N_{\rm max}$ is the number of HO shells,  and Greek and Latin
letters refer, respectively, to HO and Gamow one-body basis states. As
$\hat{V}$ appears only in HO matrix elements (\ref{V_HO_exp}), Moshinsky
transformation can be easily performed. $\hat{V}$ can also be expressed
in momentum representation in which  HO states are analytical. 
Furthermore, Gamow states appear only in overlaps $\langle a |
\alpha \rangle$. The latter ones  can be calculated without the need for
complex scaling because of the Gaussian fall-off of  HO states.

The only approximation concerns the  number  $N_{\rm max}$ of HO shells
used. Convergence for $N_{\rm max} \rightarrow +\infty$ is in fact weak,
i.e., while the energies and vectors of  resonant states properly
converge, the  matrix elements (\ref{Vosc}) do not. Note that this
scheme is very different from  HO expansion of SM states.
In the standard
SM, both kinetic and potential terms  are expressed in a HO basis, whereas
the method  of Ref.~\cite{Hag06a} invokes HO expansion  for
nuclear interaction only. Since the kinetic and
 Coulomb terms are treated exactly in the asymptotic zone, the new method is capable
 of describing weakly bound and unbound states. 

\begin{figure}[htb]
\centerline{\includegraphics[trim=0cm 0cm 0cm 0cm,width=0.7\textwidth,clip]{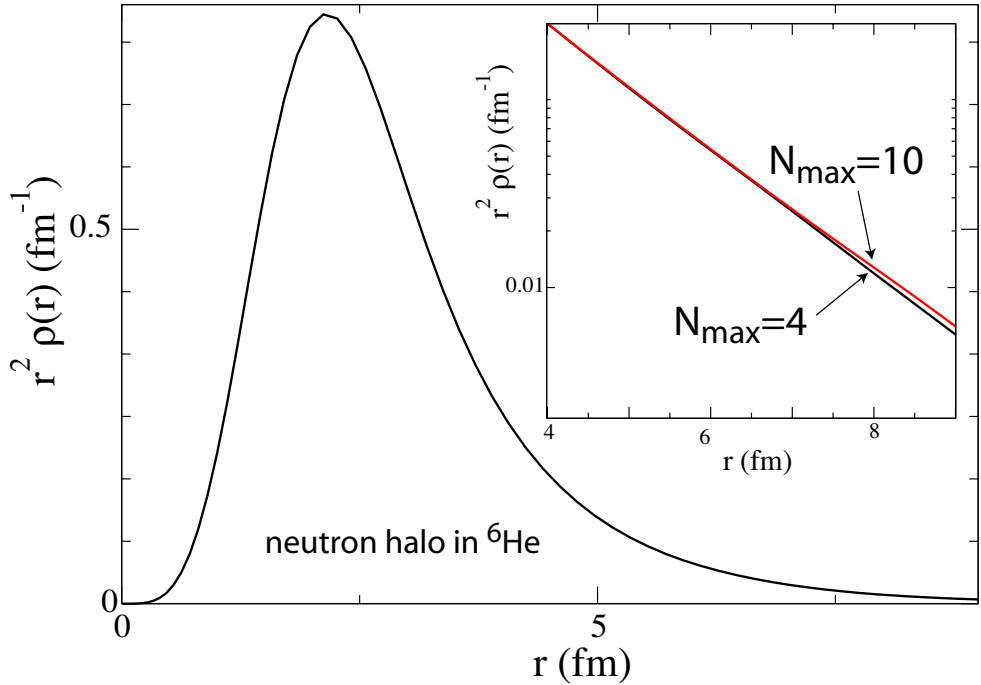}}
\caption{
Neutron halo density of the ground state of $^6$He
calculated in GSM+$V_{low-k}$~\cite{Hag06a}.
The full convergence is achieved at $N_{\rm max} \sim$10, see the inset.}
\label{He6_densities}
\end{figure}

\begin{table}[htbp]\label{Epe06_tab}
\begin{center}
\begin{tabular}{|c|c|c|c|} \hline
  $N_{\rm max}$ & $^6$He: $0_1^+$     & $^6$He: $0_2^+$   & $^{18}$O:  $4_2^+$  \\ \hline
  4         &  -0.4760 (0.0000)   & 0.9504  (-0.0467) & -1.4373  (-0.8275)  \\
  6         &  -0.4714 (0.0000)   & 0.9546  (-0.0461) & -1.4292  (-0.7600)  \\
  8         &  -0.4719  (0.0000)  & 0.9597  (-0.0453) & -1.4380  (-0.7405)  \\
  10        & -0.4721 (0.0000)    & 0.9602  (-0.0452) & -1.4400  (-0.7390)  \\
  12        & -0.4721 (0.0000)    & 0.9600  (-0.0452) & -1.4393  (-0.7401)  \\
  14        & -0.4721 (0.0000)    & 0.9601  (-0.0452) & -1.4394  (-0.7401)  \\
  16        & -0.4721 (0.0000)    & 0.9601  (-0.0453) & -1.4394  (-0.7401) \\
  18        & -0.4721 (0.0000)    & 0.9601  (-0.0453) & -1.4394  (-0.7401) \\
  20        & -0.4721  (0.0000)   & 0.9601  (-0.0453) & -1.4394  (-0.7401) \\
  \hline
\end{tabular} 
\end{center}
\caption{Energy convergence  for ${0_1}^+$ and  $ {2_1}^+ $ states in $^6$He 
and $4_2^+$  state in $^{18}$O versus the number $N_{\rm max}$ of HO
nodes in the expansion of the realistic low-momentum N$^3$LO
nucleon-nucleon interaction. The imaginary part of the energy is shown in parentheses.
A cut-off of 1.9~fm$^{-1}$ was used for the center-of-mass momentum.
The  oscillator length is $b$=2 fm. 
All energies are in MeV. (From Ref.~\cite{Hag06a}.)}
\end{table}
First applications of GSM with realistic interactions
were  carried out in Ref.~~\cite{Hag06a} with the low-momentum
potential $V_{low-k}$ \cite{Bog03} for $^6$He and $^{18}$O, viewed as
two-neutron systems. The   $^4$He or $^{16}$O cores were approximated by
a GHF potential. The same  GHF potential was  also used to generate the
Berggren basis in $spd$ neutron partial waves. $V_{low-k}$ was obtained
from the chiral N$^3$LO  interaction \cite{Epe06}  with a momentum cut-off 
$\Lambda = 1.9$ fm$^{-1}$.
The calculated energies of selected states in $^6$He and $^{18}$O
are shown in
Table~\ref{Epe06_tab} as a function of the HO states used in the expansion
(\ref{Vosc}).  One can see that the results are already stabilized at
$N_{\rm max}$$\sim$10, even for resonant states having a
sizeable width. In order to show that asymptotic properties 
can be reliably calculated in  this scheme, Fig.~\ref{He6_densities}
shows the  density distribution of the valence neutrons in the halo
nucleus $^6$He. The results are already reliable at $N_{\rm max}$=4 
while perfect convergence is attained at $N_{\rm max}$=10.
%

In the context of {\it ab initio} many-body techniques, the Berggren
ensemble has recently been employed in the coupled-cluster approach
\cite{Hag07} to study neutron-rich  He isotopes. This work demonstrates
the usefulness of a complex-energy framework for a microscopic 
description of weakly bound and unbound systems.

\subsection{Center-of-mass treatment in GSM}

In  SM approaches, special attention is paid to the removal of the
spurious center-of-mass (CM) motion \cite{Lip58,Whi77}, and GSM is no
exception. The standard SM treatment of CM is based on the Lawson  method
\cite{Law80} which, unfortunately, cannot be used within the GSM
framework due to the fact that the matrix elements of
$\mathbf{R}^2$ ($\mathbf{R}$ is the CM coordinate) 
cannot be regularized using the complex scaling technique.
One possible solution is to replace the CM Hamiltonian
by the square of the $L_z$ component of the CM
angular momentum operator $\mathbf{L} = \mathbf{R} \times \mathbf{P}$.
Using of $\mathbf{L}^2$ instead of ${L_z}^2$, would be equivalent but
more complicated because $\mathbf{L}^2$ is a four-body operator. In
this generalized Lawson method, the CM part of SM nuclear states is not
projected onto a 0s HO state, but on a $L$= 0 CM state.

Another possibility  is to solve the GSM  problem directly in relative  coordinates.
While the Jacobi coordinates have to be excluded for practical reasons, as
effects due to antisymmetrization  become quickly intractable when the number of coordinates
increases, one can define simple CM coordinates in a core+valence 
framework within the cluster-orbital shell model (COSM) \cite{Suz88}. 
In COSM, all coordinates are taken with respect to the core CM, so
that translational invariance is strictly preserved. With COSM coordinates being particle coordinates, 
 Slater determinants can be easily  defined within this scheme. The translationally-invariant COSM
 Hamiltonian can be written as:
\begin{equation}
H = H_c + \sum_{i = 1}^{A_v} \left( \frac{p_i^2}{2m} + U_i \right) +
 \sum_{i<j}^{A_v} \left( \hat{V}_{ij} + \frac{\mathbf{p_i} \cdot \mathbf{p_j}}{(A_c + 1)m} \right), \label{H_COSM}
\end{equation}
where $A_c$ and $A_v$ is  the number of core and valence particles, respectively.
The principal  difference between  the previously introduced GSM Hamiltonian
(\ref{H_def}) and that of COSM is
the term proportional to $\mathbf{p_i} \cdot \mathbf{p_j}$ which
takes into account the recoil of the active nucleons. The recoil
term    can be treated in the   momentum representation using 
the  Bessel expansion method and  the same $L^+$ complex contour
for all partial waves.
Another possibility is to expand $\mathbf{p}_i$ in a basis of bound states,
as done for realistic interactions in Sec.~\ref{GSM_real_inter}. 

Because the transformation from the laboratory to COSM frame is very
difficult to handle \cite{Katdi}, there are fundamental problems
with  using  realistic interactions in a COSM framework. 
Nevertheless, this scheme is  suitable for the development of
{\it effective} interactions for light nuclei.

\subsection{Overlap integrals and  spectroscopic factors in GSM}

Cross sections of direct reactions can be expressed as products of
two factors. The first one depends on kinetic
aspects of the reaction while the other one,
depending on the  structure 
of the states involved, is called the spectroscopic factor
\cite{Boh69,Gle63,Fro96}. The experimental deduction of
spectroscopic factors can sometimes be difficult due to probe- and
model-dependence \cite{Jen05,Han03,Gad05}. In this context,  GSM is a useful
tool as it provides both configuration mixing and a proper description of 
the particle continuum of both resonant and non-resonant character.

Spectroscopic factors are defined as the norm of the one-nucleon radial
overlap integral $u_{\ell j}(r)$ \cite{Gle63,Fro96}:
\begin{equation}
u_{\ell j}(r) = \langle \Psi^{J_{A}}_{A} | \left[ | \Psi^{J_{A-1}}_{A-1}
\rangle \otimes |\ell,j \rangle \right]^{J_A} \rangle, \label{ovf_basic_def}
\end{equation}
where $|\Psi^{J_{A}}_{A}\rangle$ and
$|\Psi^{J_{A-1}}_{A-1}\rangle$ are wave functions of nuclei $A$ and $A-1$ and
$|\ell,j\rangle$ is the angular-spin part of the channel function.
The angular-spin degrees of freedom
are integrated out in Eq.~(\ref{ovf_basic_def})
so that  $u_{\ell j}$ depends only on
the relative radial coordinate of the transferred particle.

 Using a decomposition of the $(\ell,j)$
channel in the complete Berggren basis $\mathcal{B}$, one obtains:
\begin{eqnarray}
&&u_{\ell j}(r) = \int\hspace{-1.4em}\sum_{\mathcal{B}} 
\langle \widetilde{\Psi^{J_{A}}_{A}} || a^+_{\ell j}(\mathcal{B}) || \Psi^{J_{A-1}}_{A-1} \rangle
\mbox{ } \langle r \ell j | \mathcal{B} \rangle, \label{ovf_GSM} \\
&&S^2 = \int\hspace{-1.4em}\sum_{\mathcal{B}} \langle
 \widetilde{\Psi^{J_{A}}_{A}} || a^+_{\ell j}(\mathcal{B}) ||
 \Psi^{J_{A-1}}_{A-1} \rangle^2, \label{eq3}
\end{eqnarray}
where $a^+_{\ell j} (\mathcal{B})$ is a creation operator associated
with a s.p.~Berggren
state $|\mathcal{B}\rangle$.
Since Eqs.~(\ref{ovf_GSM},\ref{eq3}) involve summation 
over all discrete Gamow states and integration over all 
scattering states along  the contour
$L_+^{\ell_j}$, the final result is 
independent of the s.p.~basis assumed. This is in contrast to standard SM
calculations where the model dependence of SFs enters through the
specific choice of a s.p.~state $a^+_{n \ell j}$. 

 Of particular interest 
is a  non-analytic behavior of direct reaction cross section near the 
particle emission threshold \cite{Wig48,Mal56,Wel63,Moo66,Abr92}.
It manifests itself by a discontinuity in the cross section
and/or of its derivatives (the so-called Wigner cusp);
it often  arises from a redistribution of
particle flux when a new reaction  channel opens. 
This phenomenon has  been theoretically studied in
Refs.~\cite{Bre57,Lan70,Hat77,Baz69a,New59,Mey63,Baz69} and
also recently in GSM \cite{Mic07,Mic07o}.

%
\begin{figure}[htb]
\centerline{\includegraphics[trim=0cm 0cm 0cm 0cm,width=0.4\textwidth,clip]
{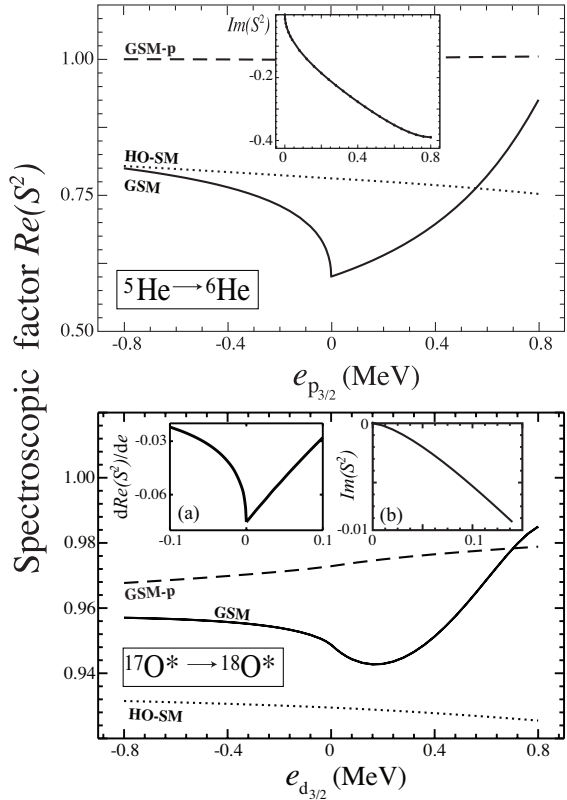}}
\caption{Top:
The real part of the overlap integral $S^2 = \langle ^6$He(g.s.)$|[^5$He(g.s.)$ \otimes p_{3/2}]^{0^+} \rangle^2$ as a
function of the energy of the $0p_{3/2}$  resonant state.
$^6$He(g.s.) is bound for all values of 
$e_{0p_{3/2}}$. The solid line
(GSM) shows the full GSM result. The dotted line (HO-SM)
corresponds to the SM calculation in the oscillator basis of
 $0p_{3/2}$ and $0p_{1/2}$ valence shells,
 and the dashed line (GSM-p) shows the GSM result
in the pole approximation.
The imaginary part of $S^2$ is shown in the inset. $S^2$ has been
normalized so as to be equal to one for vanishing residual interaction.
Bottom: Similar to the top portion 
except for the overlap 
 in the excited  $0^+_3$ state of $^{18}$O: $\displaystyle \langle
^{18}{\rm O}( 0^+_3) | [^{17}{\rm O}( 3/2_1^+) \otimes d_{3/2}]^{0^+}
\rangle^2$. The first derivative  of $S^2$  in the neighborhood of the
$e_{0d_{3/2}}$=0 threshold  is shown
in inset (a) while inset (b) displays
the imaginary part of $S^2$.}
\label{cusp}
\end{figure}
The  spectroscopic factor
[$^5$He(g.s.)$\otimes$$p_{3/2}]^{0^+}$ in $^6$He(g.s.), calculated in the GSM, is shown in
Fig.~\ref{cusp} (top) as a function of the real energy $e_{p_{3/2}}$. 
The spectroscopic
factor strongly depends on the position of the $p_{3/2}$ pole, especially
around the 1n-emission threshold in $^5$He ($e_{p_{3/2}}$=0).
As discussed in Refs.~\cite{Mic07,Mic07o}, this dependence (in particular,
a cusp due to a coupling with the
$^4$He+n+n channel)
follows the  threshold behavior of the reaction cross
section  \cite{Wig48}. Specifically, the anomalous component of the spectroscopic factor {\it
below} the 1n threshold  in $^5$He behaves as $(-e_{\ell
j})^{\ell-1/2}$.  {\it Above} the  threshold, the spectroscopic factor is complex;
the real part behaves as $(e_{\ell j})^{\ell+1/2}$ while the imaginary
part, associated with the decaying nature of $^5$He, varies as
$(e_{\ell j})^{\ell-1/2}$. 

Fig.~\ref{cusp} (bottom) illustrates the dependence of spectroscopic factors on the orbital
angular momentum $\ell$ of the transferred nucleon.  Here, we consider the
$d_{3/2}$ partial wave  in $^{18}$O and
the spectroscopic factor for the excited  $0^+_3$ state of $^{18}$O in the
channel $[^{17}{\rm O}(3/2_1^+) \otimes d_{3/2}]^{0^+}$ (see Ref.~\cite{Mic07o} for details). 
The behavior of $S^2$   is similar to that of 
$\displaystyle \langle ^6\rm{He(g.s.)} | [^5{\rm He(g.s.)} \otimes p_{3/2}]^{0^+} \rangle$,
except the variations are much weaker  and the spectroscopic factor is continuous and  smooth; it is its
derivative that exhibits a cusp  around the $0d_{3/2}$ threshold.
This is again consistent with the general expectation that, for $\ell$=2, 
$Re(S^2)$ {\em below} the 1n threshold of $^{17}$O behaves as 
$(-e_{\ell j})^{3/2}$ while {\em above} the threshold $Re(S^2)$ ($Im(S^2)$) should behave as $(e_{\ell j})^{3/2}$ ($(e_{\ell j})^{5/2}$).

To assess the role of the  continuum, both resonant and non-resonant,
the GSM results are compared in  Fig.~\ref{cusp} with the standard SM
calculations in a HO basis (HO-SM) and the results of a pole
approximation (GSM-p). In contrast to GSM,   spectroscopic factors in
HO-SM and GSM-p vary  little in the energy range considered, and no
threshold effect is seen therein. A  difference between GSM and GSM-p
results illustrates the impact of the non-resonant continuum. Clearly,
the continuum coupling should seriously be considered  in studies of
spectroscopic factors as the associated threshold effects can be as
large as those due to, e.g., short-range correlations \cite{short_range}.

In Ref.~\cite{Mic07o} one-neutron radial overlap integrals were studied
in GSM for various energy conditions of parent and  daughter nuclei.
It was concluded that 
$u_{\ell j}(r)$
can be very well
approximated by a resonant state of a one-body potential which
reproduces the complex generalized one-nucleon  separation energy 
 \begin{equation}\label{genSn}
 {\tilde S}_{1n}(N)\equiv E(N-1)-E(N) =  S_{1n}(N) -{i\over 2} \left[
 \Gamma(N-1)-\Gamma(N)
 \right].
 \end{equation}
  Indeed,
the asymptotic behavior of $u_{\ell j}(r)$ directly depends on
$\tilde{S}_{1n}$ \cite{Blo77,Tim03}:
\begin{eqnarray}
u_{\ell j}(r) \sim e^{-\kappa r} \mbox{ , } r \rightarrow +\infty \mbox{ , } 
\kappa = \sqrt{2m \tilde{S}_{1n}/ \hbar} \label{ov_asymp}.
\end{eqnarray}
 This is a straightforward generalization
of a treatment of one-nucleon overlaps often used in SM
studies  where a real optical potential is
employed with a depth adjusted to reproduce the experimental value of $S_{1n}$.

\subsection{Comparison between GSM and COSM for light nuclei}

There have been  quite a few applications of the Berggren ensemble 
to few-body systems. In the context of this review, of particular interest
are COSM calculations of Refs.~\cite{Myo01,Myo07}. They
considered a complex-scaled Hamiltonian to study light drip-line  nuclei such as $^{6-7}$He
and $^{11}$Li, described in terms of cluster+valence neutrons systems.
The  $^9$Li core of  $^{11}$Li is a complex nucleus which was described using 
 a multi-configurational wave function \cite{Myo01}. The use
of ``V" and ``T" types of coordinates in the three-body hybrid-TV model
 \cite{Ike92} (``V" being related to COSM coordinates
and ``T" to cluster coordinates) makes it possible  to take  into account
a large number of partial waves within a relatively small configuration space.

\begin{table}[htbp]\label{GSM_COSM_comparison}
\begin{center}
\begin{tabular}{|c|c|c|c|} \hline
  $(C_k)^2$         & COSM                & COSM $(\ell = 1)$ & GSM  \\ \hline
  $(0p_{3/2})^2$    & 1.211-i0.666        & 1.139-i0.742      & 1.105-i0.832 \\
  $(S1)_{p_{3/2}}$  & -0.252+i0.692       & -0.119+i0.773     & -0.060+i0.881 \\
  $(S2)_{p_{3/2}}$  & -0.042-i0.026       & -0.060-0.031      &-0.097-i0.050  \\
  sum               & 0.917               & 0.960             & 0.948 \\
                    &                     &                   &      \\
  $(0p_{1/2})^2$    & 1.447+i0.007        & 0.353-i0.077      &0.226-i0.161 \\
  $(S1)_{p_{1/2}}$  & -2.658-i0.042       & -0.534+i0.065     &-0.198+i0.224 \\
  $(S2)_{p_{1/2}}$  & 1.249+i0.034        & 0.221+i0.012      &0.025-i0.063 \\
  sum               & 0.038               & 0.040             &0.053\\
  \hline
\end{tabular} 
\end{center}
\caption{Components of the ground-state
wave function of  $^6$He in GSM \cite{Hag05} and
 COSM \cite{Mas07}. $S1$ and $S2$  indicate 
configurations with one and two particles in
the non-resonant continuum, respectively. 
The  COSM  results obtained
with partial waves up to $\ell_{\rm max}$=5 are displayed in the second
column, while 
the COSM results in a truncated space of   $p_{3/2}$ and $p_{1/2}$ partial waves
are shown in the third column. (From Ref.~\cite{Mas07}.)}
\end{table}
The COSM predictions for the He and O chains ~\cite{Mas07} have been
compared to those of GSM \cite{Mic03,Hag05}.
Table~\ref{GSM_COSM_comparison} shows the ground-state wave function
decomposition  for $^6$He  obtained in both approaches. While
configurations involving the  $0p_{3/2}$  resonant state are similar,
those associated with the  $0p_{1/2}$  broad resonant state are very
different. This is due to the large coupling to the non-resonant
continuum present in the $p_{1/2}$ partial wave with a weak
contribution of the $0p_{1/2}$ pole. While GSM is limited to   $p_{3/2}$
and $p_{1/2}$ partial waves, COSM includes all partial waves up to
$\ell$=5,  so that additional couplings change the relative occupancy of
resonant and non-resonant  $p_{1/2}$  shells. This is nicely illustrated
in a variant of COSM calculations  in which a renormalized interaction
was employed in a truncated space of   partial waves with $\ell$=1. As
seen in Table~\ref{GSM_COSM_comparison}, GSM and COSM occupations are
fairly close in this case \cite{Mas07}.

One should stress, however, that  the sum of components of the
ground-state wave function of $^6$He over resonant state and
non-resonant  continuum states for $p_{3/2}$ and $p_{1/2}$ partial waves
is practically the same in both GSM and COSM. These quantities give the
basis independent estimate of configuration mixing in the ground-state
of $^6$He as given by GSM and COSM Hamiltonians, respectively. Hence,
one can conclude that taking more partial waves does not change
quantitatively the prediction about a structure of the  ground-state
wave function of $^6$He \cite{Mic03}. 

In many aspects, GSM and COSM are complementary. The use of
complex-scaled Hamiltonian allows to calculate resonant states in COSM
in a basis of discrete bound states. Exact asymptotic behavior, which is
efficiently treated by means of  hyperspherical coordinates, has to be
ensured. In practice, however, this requirement is impossible to fulfill
with more than three clusters. This seems to be a major limitation of
COSM. The use of overlap method in GSM circumvents this problem, but the
price to pay is blurring different emission channels, so that one cannot
easily identify different partial widths. This problem is currently
under investigation.

\section{Perspectives}\label{perspectives}

Shell structure is a fundamental property of finite Fermi systems. In
nuclear physics, the nucleonic shell structure is characterized by the
appearance of large energy gaps in the s.p.~spectrum. In open shell
nuclei,  SM explains the shell evolution in terms of the configuration
mixing due to a residual interaction. Experimental discoveries over the
last decade, accompanied by theoretical  studies, have demonstrated the
fragility of the SM concept, in particular the separation of the
Hamiltonian into a  s.p.~potential (that generates shell structure with
immutable magic numbers) and a two-body residual interaction \cite{Dob07}. As a
result of configuration mixing, new magic numbers in neutron-rich nuclei
appear far from stability, while some traditional ones disappear.  Another
systematic effect of large neutron excess, illustrated  in
Fig.~\ref{OQS}, is a decrease of one-neutron separation energy  when
moving towards the neutron drip line.

GSM is the OQS formulation of SM for a {\em self-adjoint} Hamilton
operator. This choice offers a number of  conceptual advantages.
For instance,  the
transition from a bound to unbound regime, either within a single nucleus
or in the long chain of isotopes (isotones), can be viewed as an opening in
the configuration space  and described without changing the
Hamiltonian. The many-body nuclear Hamiltonian in this formulation does
not describe just one isolated nucleus $(N,Z)$, 
but all nuclei and all nuclear states that 
are coupled through various decays and captures. 
This idealization offers a right
physical picture of the many-body system and its interactions.

Already early applications of GSM to the  helium chain  revealed the
essential role of many-body continuum states in the binding mechanism of
$^6$He (bound), $^7$He (unbound), and $^8$He (bound). In addition, they
demonstrated that the configuration mixing involving continuum space may
significantly affect occupancies of different s.p.~orbits in many-body
states in the vicinity of particle-emission thresholds, thus inducing
specific variations of gaps in the spectrum of effective s.p.~energies.
The resulting picture  differs  even further from that of the standard
SM: the s.p.~shell structure  probed by GSM becomes a fragile concept in
weakly bound nuclei \cite{Dob07}. Studies of  many-body
wave functions of GSM   inform  us  how subspaces of discrete and
scattering states are  intertwined.  Having a complete basis which
allows to describe bound, weakly-bound and unbound states on the same
footing is the only way to guarantee the unitarity. This fundamental
requirement is not respected in any CQS formulation of the many-body
theory. As shown in several examples, restoration of unitarity can
strongly affect both values and behavior of  spectroscopic factors. The
unitarity lies also at the heart of the Wigner threshold effect  
in the  multichannel case. Further consequences
of the unitarity restoration on low-energy reactions, mirror
symmetry-breaking, pairing correlations, and multi-nucleon decays,
deserve further investigations.

Over the last few years, the Berggren ensemble and GSM  became an
accepted tool for nuclear structure studies. An efficient algorithm for
finding discrete resonance states in the non-resonant continuum has been
proposed \cite{Mic03} and a new method, based on the application of the
DMRG technique, was developed to tame the explosive growth of the Fock
space \cite{Rot06}. Recently, the Berggren ensemble has been applied in an {\em
ab initio} description of open quantum systems in a complex-energy
coupled-cluster approach \cite{Hag07}. The GSM serves also as a
benchmark model to test other approaches to  weakly bound
states, such as the COSM \cite{Mas07}. Progress has been made in
applying  Berggren ensembles to the mean-field description of nuclei
through Gamow-HFB.

Many challenging problems remain that could be illuminated with 
GSM ---  thanks to its ability to follow a   quantum
transition from a bound-state to unbound regime. For instance, GSM is
an excellent tool to explain an `alignment' of a many-body wave function
with a channel wave function in the vicinity of a channel threshold \cite{Ike68}. This
mechanism, which  is a likely  source of clustering effects seen near
different threshold openings, can be studied in GSM. A near-threshold
behavior  of pairing correlations is an example from  another domain
where systematic GSM studies could be most helpful. That is probably a
most transparent  illustration of a near-threshold clustering.

To improve the detailed description of bound and resonance states spectra in
GSM, a development of  effective interactions in RHS is an urgent need. 
This will open a possibility for investigations  of  configuration-mixing
effects involving continuum states in certain isotope chains,
e.g., $^9$Li (bound), $^{10}$Li (unbound), $^{11}$Li (bound), and
$^{12}$Li (unbound). In this example, GSM  could provide a quantitative
estimate of admixtures involving coupling of two neutrons (one neutron)
with different states of $^9$Li ($^{10}$Li) in the ground state of
$^{11}$Li. Recent experimental studies of the neutron transfer reaction
p($^{11}$Li,$^{9}$Li)t \cite{Tan08} provide a  strong  motivation.

Much effort has been devoted to studies of spectral  degeneracies associated
with the avoided level crossings, focusing mainly on  the
topological structure of the Hilbert space and the geometric phases
\cite{Ber84,Lau94}. Among these degeneracies, one finds  exceptional
points \cite{Kat95,Zir83,Hei91} where two eigenfunctions and associated
eigenvalues become equal. Up to now,  exceptional points have not been
studied within RHS. In this respect, GSM could provide a first realistic and
consistent many-body framework in which properties of wave functions
close to a RHS singularity could be described.

In summary,  a complex-energy shell model  opens a new window for
unification of structure and reaction aspects of weakly bound or unbound
nuclear states,  based on  the open quantum system  framework.  The
developments reviewed here are not solely limited to nuclei; they  can be of
interest in the context of other  open quantum systems
in which the coupling to the scattering continuum is present.
To describe such systems, one has to give up either the concept of the
Hilbert space or the self-adjoint nature of the Hamiltonian if one wants
to keep the simplicity and conceptual and practical 
 advantages of the standard SM framework.

\section{Acknowledgments}
Valuable discussions with  G. Hagen, A.T. Kruppa, and J. Rotureau
are gratefully acknowledged.
This work was supported in part by  the U.S. Department of Energy
under Contract Nos. DE-FG02-96ER40963 (University of Tennessee),
DE-AC05-00OR22725 with UT-Battelle, LLC (Oak Ridge National
Laboratory, and by  the Hungarian OTKA fund No. T46791.


\begin{thebibliography}{99}
\bibitem{Dob07} J. Dobaczewski, N. Michel, W. Nazarewicz, M. P{\l}oszajczak, and
J. Rotureau, {\em Prog. Part. Nucl. Phys.} 59 (2007) 432
\bibitem{[Oko03]}
J. Oko{\l}owicz, M. P{\l}oszajczak, and I. Rotter, {\em Phys. Rep.} 374 (2003) 271
\bibitem{Lane_Thomas} A.M.~Lane, and R.G.~Thomas, {\em Rev. Mod. Phys} 30 (1958) 257
\bibitem{lan} A.M. Lane,  {\em Nucl. Phys.} A {\bf 35}, 676 (1962).
\bibitem{Gel61} I.M. Gel'fand and N.Ya. Vilenkin, 
{\em Generalized Functions}, Vol. 4, Academic Press, New York (1961)
\bibitem{Mau68} K. Maurin, {\em Generalized Eigenfunction Expansions and Unitary Representations of Topological Groups}, 
Polish Scientific Publishers, Warsaw (1968)
 \bibitem{Boh78} A. Bohm, {\em The Rigged Hilbert Space and Quantum Mechanics},
 Lecture Notes in Physics 78, Springer, New York  (1978)
\bibitem{Mad05} R. de la Madrid, {\em Eur. J. Phys.}  26 (2005) 287 
\bibitem{Ludwig1} G. Ludwig, {\em Foundations of Quantum Mechanics}, 
Vol. I and II, Springer-Verlag, New York (1983)
\bibitem{Ludwig2} G. Ludwig, {\em An Axiomatic Basis of Quantum
Mechanics}, Vol. I and II, Springer-Verlag,  New York (1983)
\bibitem{Sie39} A.F.J. Siegert, {\em Phys. Rev.} 56 (1939) 750
\bibitem{Pei59} R.E. Peierls, {\em Proc. R. Soc. (London)} A 253 (1959) 16
\bibitem{Hum61} J. Humblet and L. Rosenfeld, {\em Nucl. Phys.} 26 (1961) 529
\bibitem {Lin93} P. Lind, {\em Phys. Rev.} C 47 (1993) 1903
\bibitem{Bol96} C.G. Bollini, O. Civitarese, A.L. De Paoli, and M.C. Rocca, {\em Phys. Lett.} B 382 (1996) 205
\bibitem{Fer97} L.S. Ferreira, E. Manglione, and R.J. Liotta, {\em Phys. Rev. Lett.} 78 (1997) 1640
\bibitem{Mad02} R. de la Madrid and M. Gadella, {\em Am. J. Phys.} 70 (2002) 626
\bibitem{Kap03} E. Kapu\'{s}cik and P. Szczeszek, {\em Czech. J. Phys.} 53 (2003) 1053
\bibitem{Her03} E. Hernandez, A. J\'{a}regui, and A. Mondragon, {\em Phys. Rev.} A  67 (2003) 022721
\bibitem{Kap05} E. Kapu\'{s}cik and P. Szczeszek, {\em Found. Phys. Lett.}  18 (2005) 573
\bibitem{Jul07} J. Julve, and F.J. de Urries, quant-ph/0701213.
\bibitem{Mad07} R. de la Madrid, {\em AIP Conf. Proc.} 885 (2007) 3
\bibitem{CSM_Rotter} H.W.~Barz, I.~Rotter and J.~H{\"o}hn, {\em Nucl. Phys.} A 275 (1977) 111;\\
I. Rotter, H.W. Barz, and J.~H{\"o}hn, {\em Nucl. Phys.} A 297 (1978) 237;\\
I. Rotter, {\em Rep. Prog. Phys.} 54 (1991) 635.
\bibitem{Phi77} R.J. Philpott, {\em Nucl. Phys.} A 289 (1977) 109;\\
D. Halderson and R.J. Philpott, {\em Nucl. Phys.} A 321 (1979) 295; ibid. A 359 (1981) 365.
\bibitem{SMEC} K.~Bennaceur, F.~Nowacki, J.~Oko{\l}owicz and M.~P{\l}oszajczak, {\em J. Phys.} G 24 (1998) 1631 ; 
               {\em Nucl. Phys.} A 651 (1999) 289 ; {\em Nucl. Phys.} A 671 (2000) 203 ;  \\
               R.~Shyam, K.~Bennaceur, J.~Oko{\l}owicz and M.~P{\l}oszajczak, {\em Nucl. Phys.} A 669 (2000) 65 ;\\
               K.~Bennaceur, N.~Michel, F.~Nowacki, J.~Oko{\l}owicz and M.~P{\l}oszajczak, {\em Phys. Lett.} B 488 (2000) 75;\\
   N.~Michel,   J.~Oko{\l}owicz,   F.~Nowacki    and M.~P{\l}oszajczak, {\em Nucl. Phys.} A 703 (2002) 202.     
\bibitem{SMEC_2p} J.~Rotureau, J.~Oko{\l}owicz, and M.~P{\l}oszajczak, {\em Phys. Rev. Lett.} 95 (2005) 042503 ; {\em Nucl. Phys.} A 767 (2006) 13
\bibitem{CSM_Volya} A.~Volya and V.~ Zelevinsky, {\em Phys. Rev. Lett.} 94, (2005) 052501 ; {\em Phys. Rev.} C 74, (2006) 064314
\bibitem{Bla08} B. Blank and M.~P{\l}oszajczak, {\em Rep. Prog. Phys.} 71 (2008) 046301.

\bibitem{Gam28} G.~Gamow, {\em Z. Phys.} 51 (1928) 204
\bibitem{Gur29} R.W. Gurney and E.U. Condon, {\em Phys. Rev.} 33 (1929) 127 
\bibitem{Baz69} A.I. Baz, Ya.B. Zel'dovich, and A.M. Perelomov, {\em Scattering Reactions and Decay 
in Nonrelativistic Quantum Mechanics}, Israel Program for Scientific Translations, Jerusalem (1969)
\bibitem{Ber68} T.~Berggren, \Journal{\NPA} {109}{265} {1968} 
\bibitem{Zel60} Ya.B. Zel'dovich, {\em Zh. Eksp. i Theor. Fiz.}  39 (1960) 776
\bibitem{Hok65} N. ~Hokkyo,  \Journal{\PRO} {33}{1116} {1965} 
\bibitem{Rom68} W.J. ~Romo, \Journal{\NPA} {116}{617} {1968} 
\bibitem{Zim70} J.~Zim\'anyi, M.~Zim\'anyi, B.~Gyarmati, and T.~Vertse, {\em Acta Phys. Hung.} 28 (1970) 251  
\bibitem{Zim70a} B.~Gyarmati, T.~Vertse, J. Zim\'anyi, and M. Zim\'anyi, \Journal{\PRC} {1}{1} {1970} 
\bibitem{Ban69} J. ~Bang and J.~Zim\'anyi, \Journal{\NPA} {139}{534} {1969} 
\bibitem{Gya71} B.~Gyarmati and T.~Vertse, \Journal{\NPA} {160}{523} {1971} 
\bibitem{New82} R.~Newton, {\em Scattering Theory of Waves and Particles}, Springer-Verlag, New York, Heidelberg (1982)
\bibitem{Nus72}  H.M. Nussenzveig, {\em Causality and Dispersion Relations},
  Academic Press, New York (1972)
\bibitem{Tay72} J.R. Taylor, {\em Scattering Theory}, Wiley, New York (1972)
\bibitem{Dom81} W. Domcke, {\em J. Phys.} B 14 (1981) 4889
\bibitem{Kukulin} V.I.~Kukulin, {\em Theory of Resonances}, Kluver Academic Publishers, Dordrecht, Boston, London (1989)
\bibitem{Mig72}  A.B. Migdal, A.M. Perelomov, and V.S. Popov, 
  {\em Sov. J. Nucl. Phys.}  14 (1972) 488
\bibitem{nuss} H.M. Nussenzveig,  {\em Nucl. Phys.} 11 (1959) 499
\bibitem{zavin}  R. Zavin and N. Moiseyev, {\em J. Phys.} A 37 (2004) 4619
\bibitem{Ve89} T. Vertse, P. Curutchet, R.J. Liotta, and J.~Bang,
  {\em Acta Phys. Hung.} 65 (1989) 305  
\bibitem{Bet04} R. ~Id Betan, R.J. ~Liotta, N. ~Sandulescu, and T.~Vertse, 
\Journal{\PLB} {584}{48} {2004} 
\bibitem{Bet05} R. ~Id Betan, R.J. ~Liotta, N. ~Sandulescu, and T.~Vertse, 
\Journal{\PRC} {72}{054322} {2005} 
\bibitem{Mic06} N.~Michel, W.~Nazarewicz, M.~P{\l}oszajczak, and J. ~Rotureau,
\Journal{\PRC} {74} {054305} {2006} 
\bibitem{Dun88} N.~Dunford and J.T.~Schwartz, {\em Linear Operators}, Wiley Classics Library (1988)
\bibitem{Mic04} N.~Michel, W.~Nazarewicz, and M.~P{\l}oszajczak, 
\Journal{\PRC} {70}{064313} {2004} 
\bibitem{Mic08} N.~Michel, {\em J. Math. Phys.} 49 (2008) 022109
\bibitem{Muk06} A.~Mukhamedzhanov and M.~Akin, arXiv:nucl-th/0602006
\bibitem{Muk78} N.~Mukunda, {\em Am. J. Phys.} 49 (1978) 910
\bibitem{Hei04} W.D.~Heiss, {\em J. Phys.} A  37 (2004) 2455
\bibitem{Dir58} P.A.M. Dirac, {\em The Principles of Quantum Mechanics}, Clarendon, Oxford (1958)
\bibitem{Boh97} A. Bohm, M. Gadella, and S. Maxon, {\em Computers Math. Applic.} 34 (1997) 427
\bibitem{Civ04} O. Civitarese and M. Gadella, {\em Phys. Reports} 396 (2004) 41
\bibitem{Gar76} G. Garcia-Calderon and R. Peierls, {\em Nucl. Phys.} A  265 (1976) 443
\bibitem{Zel61} Y.B. Zel'dowich, {\em JETP (Sov. Phys.)} 22 (1961) 542
\bibitem{Her84} E. Hernandez, and A. Mondragon, {\em Phys. Rev.} C 29 (1984) 722
\bibitem{Mon91} E.~Hern{\'a}ndez, A.~Mondrag{\'o}n, and
	J.~M. Vel{\'a}squez-Arcos, {\em Ann. der Physik (Leipzig)} 48 (1991) 503
\bibitem{complex_distr} M.J.~Corinthios, {\em IEE Proc.-Vis. Image Signal Process.}, 150 (2003) 69 ; 152 (2005) 97
\bibitem{sasada} K. Sasada, {\em Spectral Analysis of the Conductance of Open Quantum Dots}, PhD-thesis, University of Tokyo (2007).
\bibitem{Rom84} W.J. ~Romo, \Journal{\NPA} {419} {333} {1984} 
\bibitem{Ber78} T. ~Berggren, \Journal{\PLB} {73} {389} {1978} 
\bibitem{Ber96} T. ~Berggren, \Journal{\PLB} {373} {1} {1996} 
\bibitem{Boh89} A. Bohm, and M. Gadella, {\em Lecture Notes in Physics}, Vol. 348, 
Springer-Verlag, Berlin;
A. Bohm, {\em Int. J. Theor. Phys.} 42 (2003) 2317
\bibitem{Civ99} O. Civitarese, M. Gadella, and R. Id Betan, 
    {\em Nucl. Phys.}  A  660 (1999)  255
\bibitem{Bur96} A. B\"{u}rgers and J-M. Rost, {\em J. Phys.} B 29 (1996) 3825
\bibitem{Fri91} H. Friedrich, {\em Theoretical Atomic Physics}, Chapter 1, Springer-Verlag, New York (1991)
\bibitem{Ban78} J. ~Bang, F.A. ~Gareev, M.H. Gizzatkulov, and S.A. Goncharov, \Journal{\NPA} {309}{381} {1978} 
\bibitem{Cal86} G. ~Garcia-Calderon, \Journal{\NPA} {448}{560} {1986} 
\bibitem{Ver91} T. ~Vertse, P. ~Curutchet, R.J. ~Liotta, J.~Bang, and N.~Van
Giai, \Journal{\PLB} {264}{1} {1991} 
\bibitem{Ber93} T.~Berggren and P. ~Lind,  \Journal{\PRC} {47}{768} {1993} 
\bibitem{Alf04} T. Alferova and N. Elander,  {\em Int. J. Quantum Chem.}  97 (2004) 922
\bibitem{Rit87} M. Rittby, N. Elander, and E. Br\"andas, {\em Lecture Notes in Physics} 325 (1989) 129
\bibitem{Kry89}
P. Krylstedt, C. Carlsund, and N. Elander,  {\em J Phys. } B 22 (1989) 1051
\bibitem{Mil86} B. Milek and R. Reif, \Journal{\NPA} {458}{354} {1986} 
\bibitem{Gya79}
B. Gyarmati, A.T. Kruppa, and J. R\'evai,  \Journal{\NPA} {326}{119} {1979} 
\bibitem{Gar78}
F.A. Gareev, M.Ch. Gizzatkulov, and J. R\'evai, \Journal{\NPA} {286}{512} {1978} 
\bibitem{Ver88} T.~Vertse, P. ~Curutchet, O. ~Civitarese, L.S. ~Ferreira, 
and R.J. ~Liotta, \Journal{\PRC} {37}{876} {1988} 
\bibitem{Cur89}  P. ~Curutchet, T.~Vertse, and R.J. ~Liotta,
\Journal{\PRC} {39}{1020} {1989} 
\bibitem{Ver87} T.~Vertse, P. ~Curutchet, and R.J. ~Liotta,
{\em Lecture Notes in Physics} 325 (1989) 179
\bibitem{Dus92} G.G. ~Dussel, R.J. ~Liotta, H.~Sofia, and T.~Vertse, 
\Journal{\PRC} {46}{558} {1992} 
\bibitem{Lin94}  P. ~Lind, R.J. ~Liotta, E. ~Maglione, and T. ~Vertse, 
\Journal{\ZPA} {347}{231} {1994} 
\bibitem{Tol98} O.I. Tolstikhin, V.N. Ostrovsky, and H. Nakamura, \Journal{\PRA}  {58}{2077}{1998}
\bibitem{Toy05} K. Toyota, T. Morishita, and S. Watanable, 
\Journal{\PRC}{72} {062718}{2005}
\bibitem{Toy01} K. Toyota, T. Morishita, and S. Watanable, 
\Journal{\PRC} {63} {052721}{2001}
\bibitem{Myo98} T. Myo, A. Ohnishi, and K. Kat\=o, \Journal{\PRO} {99}{801}{1998} 
\bibitem{Hag06} G. Hagen and J.S. Vaagen, \Journal{\PRC} {73}{034321} {2006}
\bibitem{Ver82} T. ~Vertse, K.F. ~P\'al, and Z. ~Balogh, 
{\em Comput. Phys. Comm.}  27 (1982) 309 
\bibitem{Ixa84} L.Gr. ~Ixaru, {\em Numerical Methods for Differential
Equations}, Reidel, Dordrecht (1984)
\bibitem{Ixa95} L.Gr. ~Ixaru, M. ~Rizea, and T. ~Vertse, 
{\em Comput. Phys. Comm.}  85 (1995) 217 
\bibitem{Zha92} T. Zhanlav and I.V. Puzynin,
{\em Sov. J. Nucl. Phys.}  55 (1992)  349
\bibitem{Kru85} A.T. ~Kruppa and Z. ~Papp, 
{\em Comput. Phys. Comm.}  36 (1985) 59 
\bibitem{Ver95} T.~Vertse, R.J. ~Liotta, and E. ~Maglione,
\Journal{\NPA} {584}{13} {1995}
\bibitem{Blo96}  J. ~Blomqvist, O.~Civitarese, E.D. Kirchuk, R. J. ~Liotta, and T. Vertse, \Journal{\PRC} {53}{2001} {1996} 
\bibitem{Lin96} R.J. ~Liotta, E. ~Maglione, N. ~Sandulescu, and T. ~Vertse,
\Journal{\PLB} {367}{1} {1996} 
\bibitem{San97}
N. Sandulescu, O. Civitarese, R.J. Liotta, and T. Vertse, \Journal{\PRC}  {55} {1250} {1997}
\bibitem{Ver98}
T. Vertse, A. T. Kruppa, R. J. Liotta, W. Nazarewicz, N. Sandulescu, and T. R. Werner,
\Journal{\PRC} {57}{3089} {1998}
\bibitem{Ver00} 
T. Vertse, A.T. Kruppa, and W. Nazarewicz, \Journal{\PRC} {61}{064317} {2000}
\bibitem{Dus07} 
G.G. ~Dussel, R. Id ~Betan, R.J. ~Liotta, and T. ~Vertse, \Journal{\NPA} {789} {182} {2007} 
\bibitem{Dob96} J. Dobaczewski, W. Nazarewicz,
T.R. Werner, J.-F. Berger, C.R. Chinn, and J. Decharg\'e,
{\em Phys. Rev.} C 53 (1996) 2809
\bibitem{Del00} 
D.S. Delion and J. Suhonen, {\em Phys. Rev.} C 61 (2000) 024304
\bibitem{Bia01} 
A. Bianchini, R.J. Liotta, and N. Sandulescu, {\em Phys. Rev.}  C 63 (2001) 024610
\bibitem{Mag98} E.~Maglione, L.S. ~Ferreira, and R.J. Liotta, \Journal{\PRL} {81}{538} {1998} 
\bibitem{Mag99} E.~Maglione, L.S. ~Ferreira, and R.J. Liotta, 
\Journal{\PRC} {59}{R589} {1999} 
\bibitem{Ryk99} K.~Rykaczewski, et al., \Journal{\PRC} {60}{011301} {1999} 
\bibitem{Kru00}  A.T. ~Kruppa, B.~Barmore, W.~Nazarewicz, and T.~Vertse, 
\Journal{\PRL} {84}{4549} {2000} 
\bibitem{Bar00} B.~Barmore, A.T. ~Kruppa, W.~Nazarewicz, and T.~Vertse, 
\Journal{\PRC} {62}{054315} {2000} 
\bibitem{Kru04}
A.T. Kruppa and W. Nazarewicz,  {\em Phys. Rev.} C 69 (2004) 054311
\bibitem{Mar01} I. Martel, M.J.G. Borge, J. Gomez-Camacho, 
A. Poves, J. Sanchez, and O. Tengblad, {\em Nucl. Phys.} A 694 (2001) 424 
\bibitem{Gur04} 
S. A. Gurvitz, P.B. Semmes, W. Nazarewicz, and T. Vertse,
{\em Phys. Rev.} A 69 (2004) 042705
\bibitem{Bet02} R. ~Id Betan, R.J. ~Liotta, N. ~Sandulescu, and T.~Vertse, 
\Journal{\PRL} {89}{042501} {2002} 
\bibitem{Mic02} N.~Michel, W.~Nazarewicz, M.~P{\l}oszajczak, and K. Bennaceur,
\Journal{\PRL} {89}{042502} {2002} 
\bibitem{Mic03} N.~Michel, W.~Nazarewicz, M.~P{\l}oszajczak, and J.~Oko{\l}owicz,
\Journal{\PRC} {67}{054311} {2003} 
\bibitem{Bet08} R. ~Id Betan, A.T. ~Kruppa, and T.~Vertse, {\em Phys. Rev.} C, in press.
\bibitem{Rein82} W.P. Reinhardt, {\em Ann. Rev. Phys. Chem.} 33 (1982) 223
\bibitem{Agu71}  J. ~Aguilar and J. M. ~Combes, 
\Journal{\CMP} {22}{269} {1971} 
\bibitem{Bal71}  E. ~Balslev and J. M. ~Combes, \Journal{\CMP} {22}{280} {1971}
\bibitem{Sim72}  B. ~Simon, \Journal{\CMP} {27}{1} {1972}
\bibitem{Sim79} B. Simon, {\em Phys. Lett.} A 71 (1979) 211
\bibitem{gyarbor} A. ~Cs\'ot\'o, B. Gyarmati, A.T. Kruppa, K.F. P\'al,
and N. Moiseyev,  \Journal{\PRA} {41}{3469} {1990} 
\bibitem{Lef92} R. Lefebvre, \Journal{\PRA} {46}{6071} {1992}
\bibitem{Kru88} A.T. ~Kruppa, R.G.~Lovas, and B.~Gyarmati, \Journal{\PRC} {37}{383} {1988} 
\bibitem{Kru97}  A.T. ~Kruppa, P-H ~Heenen, H. ~Flocard, and R. J. ~Liotta,
\Journal{\PRL} {79}{2217} {1997} 
\bibitem{Kru01}  A. T. ~Kruppa, P-H ~Heenen, and R. J. ~Liotta,
\Journal{\PRC} {63}{044324} {2001} 
\bibitem{Nic90} C. A. ~Nicolaides, H. J. ~Gotsis, M. Chrysos, and Y. ~Komninos,
{\em Chem. Phys. Letters}  168 (1990) 570
\bibitem{Res99} T.N. Rescigno, M. Baertschy, W.A. Isaacs, and C.W. McCurdy, {\em Science} 286 (1999) 2474
\bibitem{Mez07} J.Zs. ~Mezei, A.T. ~Kruppa, and K.~Varga, 
\Journal{\FBS} {41}{233} {2007} 
\bibitem{LIs05} T. ~Li and R. ~Shakeshaft, \Journal{\PRA} {71}{052505} {2005} 
\bibitem{YHo79} Y.K. ~Ho, \Journal{\PRA} {19}{2347} {1979} 
\bibitem{Kru07} A.T. ~Kruppa, R.~Suzuki, and K.~Kat\=o, \Journal{\PRC} {75}{044602} {2007} 
\bibitem{Vau57} D.~Vautherin and M.~V{\'e}n{\'e}roni, {\em Phys. Lett.} B 25 (1957) 175
\bibitem{Hag06a} G.~Hagen, M. Hjorth-Jensen, and N.~Michel, 
\Journal{\PRC} {73}{064307} {2006} 
\bibitem{Hag05} G. Hagen, M. Hjorth-Jensen, and J.S. Vaagen,
	{\em Phys. Rev.} C 71 (2005) 044314
\bibitem{Mic09} N.~Michel, M.~Stoitsov, and K.~Matsuyanagi, in preparation
\bibitem{Bel87} S.T. Belyaev, A.V. Smirnov, S.V. Tolokonnikov, 
and S.A. Fayans, {\em Sov. J. Nucl. Phys.} 45 (1987) 783 
\bibitem{Mic05a} N.~Michel, W.~Nazarewicz, and
	M.~P{\l}oszajczak, {\em Proc. New Developments in Nuclear
	Self-Consistent Mean-Field Theories}, YITP, Kyoto, Japan,
	YITP-W-05-01, B32 (2005)
\bibitem{Bet03} R. ~Id Betan, R.J. ~Liotta, N. ~Sandulescu, and T.~Vertse, 
\Journal{\PRC} {67}{014322} {2003} 
\bibitem{Whi92} S.R. White, {\em Phys. Rev. Lett.} 69 (1992) 2363;
{\em Phys. Rev.} B 48 (1993)  10345
\bibitem{Duk04} J. Dukelsky and S. Pittel, {\em Rep. Prog. Phys.} 67 (2004) 513 
\bibitem{Duk05} U. Schollw\"ock, {\em Rev. Mod. Phys.} 77 (2005)  259 
\bibitem{Car99} E. Carlon, M. Henkel, and U. Schollw\"ock, 
{\em Eur. J. Phys.}  B 12 (1999)  99
\bibitem{Rot06} J. Rotureau, N. Michel, W. Nazarewicz, 
M. P{\l}oszajczak, and J. Dukelsky, {\em Phys. Rev. Lett.} 97 (2006)  110603
\bibitem{Rot08} J. Rotureau, N. Michel, W. Nazarewicz, M. P{\l}oszajczak, 
and J. Dukelsky, in preparation
\bibitem{moisey1} N. Moiseyev, P.R. Certain, and F. Weinhold, {\em Mol. Phys.} {\bf 36}, 1613 (1978). 
\bibitem{moisey} N. Moiseyev,  {\em Phys. Rep.} {\bf 302}, 212 (1998).
\bibitem{moisey2} N. Moiseyev, {\em Chem. Phys. Lett.} {\bf 99}, 364 (1983).
\bibitem{McC02} I. McCulloch and M. Gulacsi, {\em Europhys. Lett.} 57 (2002)  852
\bibitem{Hjo95} M.~Hjorth-Jensen, T.T.S.~Kuo, and E. Osnes,
	{\em Phys. Rep.} 261  (1995) 125
\bibitem{Suz80} K.~Suzuki and S.Y.~Lee, {\em Prog. Theor. Phys.} 
64 (1980) 2091;
K.~Suzuki, {\em Prog. Theor. Phys.} 68 (1982) 246; 
K.~Suzuki, and R.~Okamoto, {\em Prog. Theor. Phys.} 93 (1995) 905
\bibitem{Bog03} S.~Bogner, T.T.S.~Kuo, and A. Schwenk, {\em Phys. Rep.} 386 (2003) 1
\bibitem{Wir95} R.~Wiringa, V.G.J.~Stoks, and R.~Schiavilla, \Journal{\PRC} {51}{38} {1995} 
\bibitem{Mac01} R.~Machleidt, \Journal{\PRC} {63}{024001} {2001} 
\bibitem{Epe06} E.~Epelbaum, {\em Prog. Part. Nucl. Phys.} 57 (2006) 654
\bibitem{Law80} R.D.~Lawson, {\em Theory of the Nuclear Shell Model}, Clarendon-Press, Oxford (1980)
\bibitem{Bal69} R.~Balian and E.~Brezin, {\em Nuovo Cim.} 61 (1969) 403
\bibitem{Won72} C.W.~Wong and D.M.~Clement, \Journal{\NPA} {183}{210} {1972} 
\bibitem{Kun79} C.L.~Kung, T.T.S.~Kuo, and K.F.~Ratcliff, \Journal{\PRC} {19}{1063} {1979} 
\bibitem{Hag07} G.~Hagen, D.J.~Dean, M.~Hjorth-Jensen, and
	T.~Papenbrock, {\em Phys. Lett.} B 656 (2007) 169
\bibitem{Lip58} H.J.~Lipkin, {\em Phys. Rev.} 110 (1958) 1395
\bibitem{Whi77} R.R.~Whitehead, A.~Watt, B.J.~Cole, and I.~Morrison,
	{\em Adv. Nucl. Phys.} 9 (1977) 123
\bibitem{Suz88} Y.~Suzuki and K.~Ikeda, {\em Phys. Rev.} C 38 (1988) 410
\bibitem{Katdi} K.~Kat\=o, private communication
\bibitem{Boh69} A. Bohr and B.R. Mottelson, {\em Nuclear Structure}, Vol. 1, W.A. Benjamin, New York (1969)
\bibitem{Gle63} N.K. Glendenning, {\em Ann. Rev. Nucl. Sci.} 13 (1963) 191; 
         N.K. Glendenning, {\em Direct Nuclear Reactions}, Academic Press Inc. (1983)
\bibitem{Fro96} P. Fr{\"o}brich and R. Lipperheide, {\em Theory of Nuclear Reactions}, Oxford 
Science Publications, Clarendon Press, Oxford (1996)
\bibitem{Jen05} M.B. Tsang, Jenny Lee, and W.G. Lynch, {\em Phys. Rev. Lett.} {95} (2005) {222501}
\bibitem{Han03}  P.G. Hansen and J.A. Tostevin, {\em Ann. Rev. Nucl. Part. Sci.} 53 (2003) 219
\bibitem{Gad05} A. Gade {\em et al.},  {\em Eur. Phys. J.} A  25 (2005) s01, 251;
D. Bazin {\em et al.}, {\em Phys. Rev. Lett.}  {91} (2003) {012501}
\bibitem{Wig48} E.P. Wigner,  {\em Phys. Rev.}  {73} (1948) {1002}
\bibitem{Mal56} P.R. Malmberg, {\em Phys. Rev.}  {101} (1956) {114}
\bibitem{Wel63} J.T. Wells, A.B. Tucker, and W.E. Meyerhof, \Journal{\PREV} {131} {1644} {1963} 
\bibitem{Moo66} C.F. Moore {\em et al.},  \Journal{\PRL}{17}{926} {1966}
\bibitem{Abr92} S. Abramovich, B. Guzhovskij, and L. Lasarev, {\em Sov. J. Part. Nucl.} 23 (1992) 129
\bibitem{Bre57} G. Breit  \Journal{\PREV}{107}{1612}{1957}
\bibitem{Lan70}  A.M. Lane,  \Journal{\PLB}{33}{274}{1970} 
\bibitem{Hat77} C.~Hategan, {\em Ann. Phys.}  116 (1978) 77
\bibitem{Baz69a} A.I. Baz, {\em JETP (Sov. Phys.)} 6 (1957) 709
\bibitem{New59} R.G. Newton, \Journal{\PREV}{114}{1611}{1959} 
\bibitem{Mey63}  W.E. Meyerhof, \Journal{\PREV}{129}{692}{1963} 
\bibitem{Mic07} N. Michel, W. Nazarewicz, and M. P{\l}oszajczak, {\em Phys. Rev.} C 75 (2007) 031301(R)
\bibitem{Mic07o} N. Michel, W. Nazarewicz, and M. P{\l}oszajczak, {\em Nucl. Phys.} A 794 (2007) 29
\bibitem{short_range} W.H.~Dickhoff, and C.~Barbieri, {\em Prog. Part. Nucl. Phys.} 52 (2004) 377
\bibitem{Blo77} L.D. Blokhintsev, I. Borbely, and E.I. Dolinskii, {\em Sov. J. Part. Nucl.} 8 (1977) c485
\bibitem{Tim03} N.K. Timofeyuk, L.D. Blokhintsev, and J.A. Tostevin, {\em Phys. Rev.} C {68} (2003) {021601(R)}
\bibitem{Myo01} T. Myo, K. Kat\=o, S. Aoyama, and K. Ikeda, {\em Phys. Rev.} C  63 
(2001) 054313;
T. Myo, K. Kat\=o, S. Aoyama, and K. Ikeda, {\em Phys. Lett.} B  576 (2003) 281
\bibitem{Myo07} T.~Myo, K.~Kat\=o, and K.~Ikeda, {\em  Phys. Rev.} C 76 (2007) 054309
\bibitem{Ike92} K.~Ikeda, {\em Nucl. Phys.} A 538 (1992) 355c; 
S.~Mukai, S.~Aoyama, K.~Kat\=o, and K.~Ikeda, {\em Prog. Theor. Phys.} 99 (1998) 381; 
Y.~Tosaka, Y.~Suzuki, and K.~Ikeda, {\em Prog. Theor. Phys.} 83 (1990) 1140
\bibitem{Mas07} H.~Masui, K.~Kat\=o, and K.~Ikeda, {\em Phys. Rev.} C 75 (2007) 034316
\bibitem{Ike68} K. Ikeda, N. Takigawa, and. H. Horiuchi,
     {\em Prog. Theor. Phys. Suppl. Extra Number}, 464 (1968)
\bibitem{Tan08} I. Tanihata et al., arXiv:0802.1778
\bibitem{Ber84} M.V. Berry, {\em Proc. R. Soc. London}, Ser. A 392 (1984) 45
\bibitem{Lau94} H.-M. Lauber, P. Weidenhammer, and D. Dubbers, {\em Phys. Rev. Lett.} 72 (1994) 1004;
D.E. Manolopoulos and M.S. Child, {\em Phys. Rev. Lett.} 82 (1999) 2223;
 F. Pistolesi and N. Manini. {\em Phys. Rev. Lett.} 85 (2000) 1585;
 C. Dembowski {\em et al.}, {\em Phys. Rev. Lett.} 86 (2001) 787
\bibitem{Kat95} T. Kato, {\em Perturbation Theory for Linear Operators}, Springer Verlag, Berlin (1995)
\bibitem{Zir83} M.R. Zirnbauer, J.J.M. Verbaarschot, and H.A. Weidenm\"{u}ller, {\em Nucl. Phys.} A  411 (1983) 161
\bibitem{Hei91} W.D. Heiss and W.-H. Steeb, {\em J. Math. Phys.} 32 (1991) 3003



\end{thebibliography}
\end{document}